\documentclass[aps,prb,showpacs,superscriptaddress]{revtex4}
\usepackage{times}
\usepackage{epsfig}
\usepackage{graphicx}
\usepackage{amsmath}
\usepackage{mathptmx}
\usepackage{amssymb}
\usepackage{color}

\begin{document}

\title{Intravalley Multiple Scattering of Quasiparticles in Graphene}

\author{J. Y. Vaishnav}
\affiliation{Department of Physics, Bucknell University, Lewisburg, Pennsylvania 17837,USA}
%\author{Charles W. Clark}
%\affiliation{Joint Quantum Institute, National Institute of Standards and Technology,
%Gaithersburg, Maryland 20899,USA}
\author{Justin Q. Anderson}
%If no objections I'd like to list both CSM and NIST, as I did work for this paper from both institutions. -JQA
%Listing NIST forces the paper to pass through NIST internal review, which will delay it
%for another few months.  Since Charles is not an author, we elected to move NIST to the acknowledgments.  -JV
\affiliation{Department of Physics, Colorado School of Mines, Golden, Colorado 80401,USA}
\author{Jamie D. Walls}
\email{Corresponding author: jwalls@miami.edu}
\affiliation{Department of Chemistry,
 University of Miami, Coral Gables, Florida 33124,USA}

\begin{abstract}
We develop a theoretical description of intravalley scattering of quasiparticles in graphene from multiple short-range scatterers of size much greater than the carbon-carbon bond length.  Our theory provides a method to rapidly calculate the Green's function in graphene for arbitrary configurations of scatterers.  We demonstrate that non-collinear multiple scattering trajectories generate pseudospin rotations that alter quasiparticle interference, resulting in significant modifications to the shape, intensity, and pattern of the interference fringes in the local density of states (LDOS).  We illustrate these effects via theoretical calculations of the LDOS for a variety of scattering configurations in single layer graphene.  A clear understanding of impurity scattering in graphene is a step towards exploiting graphene's unique properties to build future devices.
\end{abstract}
\maketitle

\section{Introduction}

  Due to the fact that the graphene unit cell in single layer graphene consists of two inequivalent carbon atoms [the $A$ and $B$ carbon atoms shown in Fig. \ref{fig:pwave}(A)], the quasiparticles in graphene can be described as a pseudospin.  This imbues the quasiparticles in graphene with a sense of chirality that depends upon the pseudospin direction.   Compared to conventional two-dimensional electron gases (2DEGs), graphene's chiral nature provides it with tremendous potential for use in ultrafast electronic and pseudospintronic devices\cite{Neto09,Abergel10}.   An important question that has been extensively studied is the effect of point defects or point scatterers upon the local density of states (LDOS) in graphene, where the size of the ``point" scatterer is on the order of or smaller than the $C-C$ bond, $b=0.142$ nm\cite{Mizes89,Wehling07,Peres07,Bena08,Bena09a,Basko08}.  Scattering from defects of this size gives rise to both intravalley and intervalley scattering, the latter of which significantly affects the transport properties in graphene.  Most studies have focused on calculating the effects of scattering from one or two point defects (a quantum corral of point defects in graphene has also been studied\cite{Wang07}) on the LDOS and have used the scattering amplitudes from single point defects as input in modeling the mobilities in graphene\cite{Fang08}.  In these models, the calculated LDOS is sensitive to the particular placement of the scatterer(s) on the graphene lattice.  However, as the size or scattering length of the scatterers becomes much larger than the $C-C$ bond length, the LDOS should become less sensitive to the actual placement of the scatterers on the graphene lattice and should depend only on the relative configuration of scatterers especially at low wavelengths $k\ll \frac{2\pi}{b}$.  While controlling the distribution of point defects in a device is challenging, the controlled introduction and arrangement of large scatterers in graphene should be easier to accomplish experimentally by, for instance, placing small metallic islands placed atop or below a graphene surface\cite{Titov10,Kessler10}.  In this case, such large scatterers should predominately cause {\it intravalley} scattering.

  In this work, we present a theory for intravalley scattering of quasiparticles from multiple scatterers when the wavelength of the quasiparticles is comparable to the size of the scatterers. The $t-$matrix operators for all partial waves for a single scatterer are derived and used in developing a theory of multiple scattering in graphene. Our calculations suggest many partial waves must be included in order to accurately calculate the LDOS even when the quasiparticle's wavelength exceeds the scattering length.  That is, the usual $s$-wave approximation, which is ubiquitous in low-energy scattering of free particles, breaks down in graphene.   We calculate both the Green's function and the LDOS in graphene in the presence of multiple scatterers.  From our calculations, the chiral or pseudospin nature of graphene results in changes to the intensity, shape, and pattern of quasiparticle interference in the LDOS compared to the LDOS found in a 2DEG.  Our calculations should be experimentally verifiable using scanning tunneling microscopy (STM) or in ``simulations" of graphene systems, such as in the scattering of an atom from multiple atoms confined in a hexagonal optical lattice \cite{hexagonatom}.

%As graphene has the potential to play a tremendous role in electronics, understanding the role that interference between the chiral or pseudospin nature of graphene, and its effect on %graphene's transport properties may eventually allow us to exploit that interference--perhaps by using an STM to selectively add impurities to the graphene in order to control the current %through it.
\section{Theory}
\subsection{Graphene preliminaries}
In this section, we briefly review the physics of graphene\cite{Geim07} that will be relevant to the scattering theory subsequently developed in this paper.  A single layer of graphene  consists of two displaced triangular lattices of carbon atoms, $A$ and $B$, that generate a hexagonal or honeycomb structure as shown in Fig.~\ref{fig:pwave}(A).  Each unit cell of graphene consists of an $A$ and $B$ lattice site where the $j^{th}$ unit cell is defined by two integers, $j\equiv[m,n]$, and the positions of the $A$ and $B$ lattice sites in the $j^{th}$ unit cell are $\vec{r}_{j}^{A}=m\vec{a}_{+}+n\vec{a}_{-}$ and $\vec{r}_{j}^{B}=\vec{r}^{A}_{j}-b\widehat{y}$  respectively, where $\vec{a}_{\pm}=\pm\frac{\sqrt{3}b}{2}\widehat{x}+\frac{3b}{2}\widehat{y}$ are the lattice vectors for the honeycomb lattice and $b=1.42$~\AA~ is the carbon-carbon bond length.   The corresponding reciprocal lattice vectors are given by $\vec{a}_{\pm}^{*}=\frac{2\pi}{3b}\left(\pm\sqrt{3}\widehat{x}+\widehat{y}\right)$, with $\vec{a}_{\pm}^{*}\cdot\vec{a}_{\pm}=2\pi$ and $\vec{a}_{\pm}^{*}\cdot\vec{a}_{\mp}=0$.

    Defining the following lattice wave functions over $N_{lat}$ unit cells\cite{Bena09}:
    \begin{eqnarray}
    \Phi^{\text{lat}}_{A}(\vec{r},\vec{k})&=&\frac{1}{\sqrt{N_{lat}}}\sum_{j}e^{ i\vec{k}\cdot\vec{r}^{A}_{j}}\phi_{Aj}(|\vec{r}-\vec{r}^{A}_{j}|)\nonumber\\
    \Phi^{\text{lat}}_{B}(\vec{r},\vec{k})&=&\frac{1}{\sqrt{N_{lat}}}\sum_{j}e^{i\vec{k}\cdot\vec{r}^{A}_{j}}\phi_{Bj}(|\vec{r}-\vec{r}^{B}_{j}|)
    \label{eq:latticewave}
      \end{eqnarray}
      where $\phi_{A(B)j}(|\vec{r}-\vec{r}^{A(B)}_{j}|)$ denotes an orbital centered on the A(B) lattice site in the $j^{th}$ unit cell.    Tight-binding calculations\cite{Wallace47} taking into account nearest neighbor coupling have previously demonstrated that there exist four distinct, zero-energy states in graphene that are given by $\frac{1}{\sqrt{2}}\left(\Phi^{\text{lat}}_{A}(\vec{r},\vec{K})\pm\Phi^{\text{lat}}_{B}(\vec{r},\vec{K})\right)$ and  $\frac{1}{\sqrt{2}}\left(\Phi^{\text{lat}}_{A}(\vec{r},-\vec{K})\pm\Phi^{\text{lat}}_{B}(\vec{r},\vec{-K})\right)$, where $\pm\vec{K}=\pm\frac{\vec{a}^{*}_{+}-\vec{a}^{*}_{-}}{3}=\pm\frac{4\pi\sqrt{3}}{9b}\widehat{x}$.   The dispersion relation for graphene is linear when expanded about $\pm\vec{K}$ for small $\vec{k}$ [$|\vec{k}|\ll|\vec{K}|$].  In this case, the lattice wave functions at $\vec{k}\pm\vec{K}$ can be written in terms of the lattice wave functions at $\pm\vec{K}$,  $\Phi^{\text{lat}}_{A(B)}(\vec{r},\pm\vec{K}+\vec{k})\approx e^{i\vec{k}\cdot\vec{r}}\Phi^{\text{lat}}_{A(B)}(\vec{r},\pm\vec{K})$.  The energies and eigenstates of the tight-binding Hamiltonian about $\pm\vec{K}$ can be written as $E_{\pm\vec{K}}$ and $\Psi^{\text{clean}}_{\pm\vec{K}}(\vec{r},\vec{k})=c^{\pm}_{A}(\vec{r},\vec{k})\Phi_{A}(\vec{r},\pm \vec{K})+c^{\pm}_{B}(\vec{r},\vec{k})\Phi_{B}(\vec{r},\pm\vec{K})$ respectively, where the coefficients or ``envelope" functions, $c^{\pm}_{A(B)}(\vec{r},\vec{k})$, and energies are determined by solving the following equations:% following equations:
%to the linearized Hamiltonian $\widehat{H}_{\pm \vec{K}}$:
% \begin{equation}
%H_{\pm\vec{K}}=\mp i\hbar v_{F}\left[\begin{array}{cc}
%0 & e^{-i\theta}(\partial_{r}-\frac{i}{r}\partial_{\theta})\\
%e^{i\theta}(\partial_{r}+\frac{i}{r}\partial_{\theta}) & 0\end{array}\right]_{\pm \vec{K}}.\label{eq:h}\end{equation}
\begin{eqnarray}
E_{\pm\vec{K}}c^{\pm}_{A}(\vec{r},\vec{k})&=&\mp i\hbar\nu_{F}e^{-i\theta}\left(\frac{\partial}{\partial r}-\frac{i}{r}\frac{\partial}{\partial\theta}\right)c^{\pm}_{B}(\vec{r},\vec{k})=\pm \hbar\nu_{F}k\hat{L}_{-}\left[c_{B}^{\pm}(\vec{r},\vec{k})\right]\nonumber\\
E_{\pm\vec{K}}c^{\pm}_{B}(\vec{r},\vec{k})&=&\mp i\hbar\nu_{F}e^{i\theta}\left(\frac{\partial}{\partial r}+\frac{i}{r}\frac{\partial}{\partial\theta}\right)c^{\pm}_{A}(\vec{r},\vec{k})=\pm\hbar\nu_{F}k\hat{L}_{+}\left[c_{A}^{\pm}(\vec{r},\vec{k})\right]\nonumber\\
\label{eq:h}\end{eqnarray}
 where $\hbar\nu_{F}=1.0558\times 10^{-28}$J-m, and $\hat{L}_{\pm}=\frac{1}{ik}e^{\pm i\theta}\left(\frac{\partial}{\partial_{r}}\pm\frac{i}{r}\frac{\partial}{\partial_{\theta}}\right)=\frac{1}{ik}\left(\frac{\partial}{\partial_{x}}\pm i\frac{\partial}{\partial_{y}}\right)$.  The solutions to Eq.~(\ref{eq:h}) are parameterized by the wave vector $\vec{k}=(k\cos\theta_{\vec{k}},k\sin\theta_{\vec{k}})$ with $|\vec{k}|\ll|\vec{K}|$ and are given by\begin{eqnarray}
\Psi^{\text{clean}}_{\vec{K}}(\vec{r},\vec{k},\pm)&=&\frac{1}{\sqrt{2}}e^{i\vec{k}\cdot\vec{r}}\left(\begin{array}{c}1\\
\pm e^{i\theta_{\vec{k}}}\end{array}\right)_{\vec{K}}=e^{i\vec{k}\cdot\vec{r}}|\theta_{\vec{k}},\pm\rangle_{\vec{K}}\equiv\frac{e^{i\vec{k}\cdot\vec{r}}}{\sqrt{2}}\left(\Phi_{A}(\vec{r},\vec{K})\pm e^{i\theta_{\vec{k}}}\Phi_{B}(\vec{r},\vec{K})\right)\nonumber\\
\Psi^{\text{clean}}_{-\vec{K}}(\vec{r},\vec{k},\pm)&=&\frac{1}{\sqrt{2}}e^{i\vec{k}\cdot\vec{r}}\left(\begin{array}{c}1\\
\pm e^{i\theta_{\vec{k}}}\end{array}\right)_{-\vec{K}}=e^{i\vec{k}\cdot\vec{r}}|\theta_{\vec{k}},\pm\rangle_{-\vec{K}}\equiv\frac{e^{i\vec{k}\cdot\vec{r}}}{\sqrt{2}}\left(\Phi_{A}(\vec{r},-\vec{K})\pm e^{i\theta_{\vec{k}}}\Phi_{B}(\vec{r},-\vec{K})\right)
\label{eq:planewaveeig}\end{eqnarray}
where the energies are $E_{\vec{K}}=\pm \hbar v_{F}k$ for $\Psi^{\text{clean}}_{\vec{K}}(\vec{r},\vec{k},\pm)$ and $E_{-\vec{K}}=\mp \hbar v_{F} k$ for $\Psi^{\text{clean}}_{-\vec{K}}(\vec{r},\vec{k},\pm)$ [the superscript ``clean" refers to graphene in the absence of scatterers].  The linear dispersion relation about $\pm\vec{K}$ gives the same results as the exact tight-binding calculations for $|\vec{k}|b\leq 0.2$.  Such a linear dispersion relation is analogous to the dispersion relation found for massless Dirac fermions\cite{Semenoff84,Novoselov05}, and therefore the wave vectors, $\pm\vec{K}$, are often referred to as Dirac points.

From Eq.~(\ref{eq:planewaveeig}), the envelope functions for $\Psi^{\text{clean}}_{\pm\vec{K}}$ are plane waves.  Since we are developing a theory for scattering from localized scatterers, we can express the eigenstates for the graphene Hamiltonian in cylindrical coordinates, which are given by:
\begin{equation}
\mathcal{H}_{l,\vec{K}}^{(1,2)}(\vec{r},k,\pm)e^{il\theta}=\frac{1}{\sqrt{2k}}\left[\begin{array}{c}
H_{l}^{(1,2)}(kr)e^{il\theta}\\
\pm iH_{l+1}^{(1,2)}(kr)e^{i(l+1)\theta}\end{array}\right]_{\vec{K}}.\label{eq:hankelvec}\end{equation} about the $\vec{K}$ Dirac point
[the cylindrical eigenstates about the $-\vec{K}$ Dirac point are found by simply replacing $\vec{K}$ by $-\vec{K}$ in Eq.~ (\ref{eq:hankelvec})], and $H_{l}^{(1,2)}(kr)$ are Hankel functions of order $l$.  The states $\mathcal{H}^{(1)}_{l,\pm\vec{K}}e^{il\theta}$ and $\mathcal{H}^{(2)}_{l,\pm\vec{K}}e^{il\theta}$ represent outgoing and incoming cylindrical waves about $\vec{r}=0$, respectively.
\subsection{The $t-$matrix for intravalley scattering}
In the following,
a theory for intravalley scattering $\left[\vec{k}\pm\vec{K}\rightarrow \vec{k}'\pm\vec{K}\right]$ from cylindrically symmetric scatterers placed atop
a graphene sheet will be developed.  First, we consider the case of a single, cylindrically symmetrical scatterer of radius $a$ centered at $\vec{r}_{n}$ and represented by a potential acting equally on both the $A$ and $B$ lattice sites\cite{Cheianov07}, $V(\vec{r})=V_{0}$ for $|\vec{r}-\vec{r}_{n}|\leq a$ and $V(\vec{r})=0$ for $|\vec{r}-\vec{r}_{n}|>a$;  that is, the scatterer behaves as a uniform potential of radius $|\vec{r}-\vec{r}_{n}|\leq a$.  Scattering from such potentials in graphene has been previously studied\cite{novikov07,Katsnelson07,Hentschel07}. For intervalley scattering processes $\left[\vec{k}\pm\vec{K}\rightarrow\vec{k}'\mp\vec{K}\right]$ to be neglected, we require
$\left|\int^{2\pi}_0\int^{a}_{0}re^{i2\vec{K}\cdot\vec{r}}dr d \theta \right|\ll\pi a^{2}$, which implies that the radius of the scatterer should satisfy $a\gg\frac{2\pi}{|\vec{K}|}=3.7$~\AA.
\begin{figure}
\begin{centering}
\includegraphics[scale=0.25]{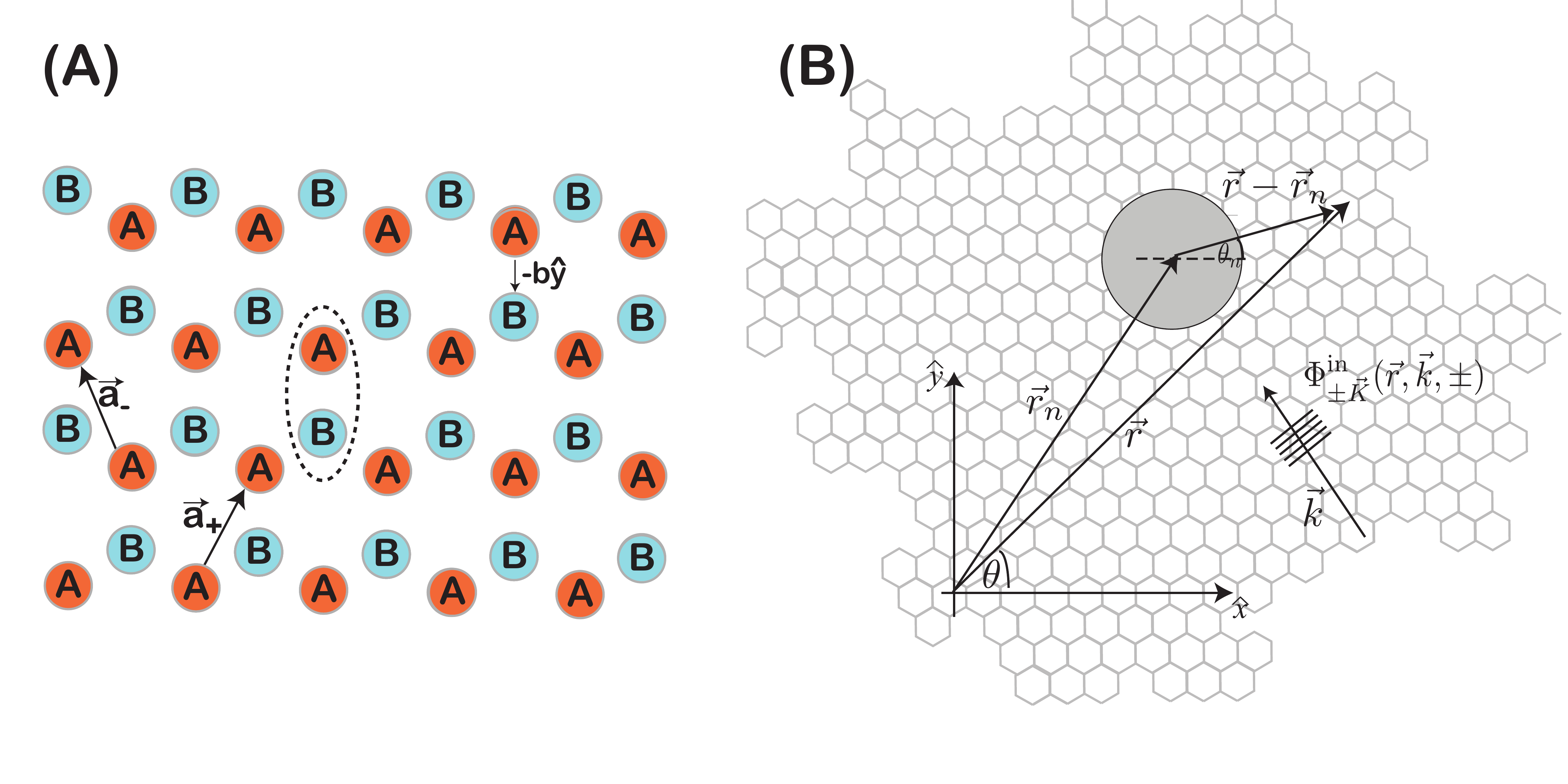}\caption{[Color Online](A) Honeycomb lattice structure of monolayer graphene consisting of two triangular lattices [denoted by $A$ (red) and $B$ (blue) carbon atoms] that are shifted relative to each other by $-b\hat{y}$ where $b=1.42$~\AA~ is the $C-C$ bond length.  Each unit cell consists of an $A$ and $B$ lattice site [a typical unit cell is denoted by the dotted ellipse in Fig.~\ref{fig:pwave}(A)], with the position of the $j^{th}$ unit cell given by $\vec{r}^{A}_{j}=m\vec{a}_{+}+n\vec{a}_{-}$ where $j\equiv[m,n]$ and $\vec{a}_{\pm}=\pm\frac{\sqrt{3}b}{2}\widehat{x}+\frac{3b}{2}\widehat{y}$ are the lattice vectors.  (B) Scattering configuration in graphene for an incoming ``wave", $\Phi^{in}_{\pm\vec{K}}(\vec{r},\vec{k},\pm)=e^{i\vec{k}\cdot\vec{r}}|\theta_{\vec{k}}\rangle_{\pm\vec{K}}$ incident upon a scatterer located at $\vec{r}_{n}$.  \label{fig:pwave}}
\par\end{centering}
\end{figure}

We will first consider intravalley scattering of a free particle wave with $E\geq 0$ about each Dirac point, $\pm\vec{K}$, $\Phi^{in}_{\pm\vec{K}}(\vec{r},\vec{k},\pm)=\Psi^{\text{clean}}_{\pm\vec{K}}(\vec{r},\vec{k},\pm)$, incident upon a scatterer located at $\vec{r}_{n}$ [Fig.~\ref{fig:pwave}(B)].
%Using the Jacobi-Anger expansion, and defining\begin{equation}
%\mathcal{J}_{l,\pm}(\vec{\rho}_{n},\vec{k})=\frac{1}{\sqrt{2k}}\left(\begin{array}{c}
%J_{l}(k\rho_{n})\\
%\pm iJ_{l+1}(k\rho_{n})e^{i\theta_{n}}\end{array}\right)\label{eq:besselvec}\end{equation}
% the incident wave is \begin{eqnarray}
%\Phi^{\pm}(\vec{r}) & = & %\sqrt{k}e^{i\vec{k}\cdot\vec{r}_{n}}\sum_{l}i^{l}\mathcal{J}_{l,\pm}(\vec{\rho}_{n},\vec{k})e^{il(\theta_{n}-\theta_{\vec{k}})}\nonumber \\
% & = & \sqrt{k}e^{i\vec{k}\cdot\vec{r}_{n}}\sum_{l}i^{l}\left(\frac{1}{2}\mathcal{H}_{l,\pm}^{(1)}(\vec{\rho}_{n},\vec{k})+\frac{1}{2}\mathcal{H}_{l,\pm}^{(2)}(\vec{\rho}_{n},\vec{k})\right)e^{il(\theta_{n}-\theta_{\vec{k}})}.\label{eq:ja}\end{eqnarray}
In this case, the full scattered wavefunction can be written as follows:\[
\Psi_{\pm\vec{K}}(\vec{r},\vec{k},\pm)=\sqrt{k}e^{i\vec{k}\cdot\vec{r}_{n}}\sum_{l}i^{l}\left(\frac{1}{2}e^{2i\delta_{l}}\mathcal{H}_{l,\pm\vec{K}}^{(1)}(\vec{\rho}_{n},\vec{k},\pm)+\frac{1}{2}\mathcal{H}_{l,
\pm\vec{K}}^{(2)}(\vec{\rho}_{n},\vec{k},\pm)\right)e^{il(\theta_{n}-\theta_{\vec{k}})}.\]
where $\rho_{n}=|\vec{r}-\vec{r}_{n}|$, $e^{i l\theta_{n}}=\left[\frac{(\vec{r}-\vec{r}_{n})\cdot\widehat{x}+ i(\vec{r}-\vec{r}_{n})\cdot\widehat{y}}{\rho_{n}}\right]^{l}$, and $\delta_{l}$ are phase shifts of the outgoing cylindrical partial waves, $\mathcal{H}^{(1)}_{l,\pm\vec{K}}$. The scattered wave function is given by: \begin{eqnarray}
\Psi^{s}_{\pm\vec{K}}(\vec{r},\vec{k},\pm) & = & \Psi_{\pm\vec{K}}(\vec{r},\vec{k},\pm)-\Phi^{in}_{\pm\vec{K}}(\vec{r},\vec{k},\pm)\label{eq:scatteredwave}\\
 & = & \sqrt{k}e^{i\vec{k}\cdot\vec{r}_{n}}\sum_{l}i^{l}s_{l}\mathcal{H}_{l,\pm\vec{K}}^{(1)}(\vec{r}-\vec{r}_{n},\vec{k},\pm)e^{il(\theta_{n}-\theta_{\vec{k}})}\nonumber \end{eqnarray}
where $s_{l}=\frac{e^{2i\delta_{l}}-1}{2}$ is the scattering amplitude of the $l^{th}$ partial wave, which is determined by continuity of the wavefunction at $\rho_{n}=a$ and is given by
\begin{equation}
s_{l}=\frac{J_{l}(k^{\prime}a)J_{l+1}(ka)-J_{l}(ka)J_{l+1}(k^{\prime}a)}{J_{l+1}(k^{\prime}a)H_{l}^{\text{(1)}}(ka)-J_{l}(k^{\prime}a)H_{l+1}^{\text{(1)}}(ka)},\label{eq:phaseshift}
\end{equation}
where $k'=k-\frac{V_{0}}{\hbar v_{F}}$ and $J_l$ is a bessel function of order $l$. Note that $s_{l}$ is the same for scattering about $\vec{K}$ and $-\vec{K}$ and satisfies the unitarity condition $-\text{Re}[s_{l}]=|s_{l}|^{2}$ for all $l$.

%\textcolor{black}{The $l^{{\rm th}}$ partial wave of the wavefunction
%inside the well is\[
%\Psi_{l,<}^{\pm}\propto\left(\begin{array}{c}
%J_{l}(k^{\prime}\rho_{n})\\
%\pm iJ_{l+1}(k^{\prime}\rho_{n})e^{i\theta_{n}}\end{array}\right).\]
%The $l^{{\rm th}}$ partial wave of the wavefunction outside the well
%is\[
%\Psi_{l,>}^{\pm}\propto e^{2i\delta_{l}}\left(\begin{array}{c}
%H_{l}^{(1)}(k\rho_{n})\\
%\pm iH_{l+1}^{(1)}(k\rho_{n})e^{i\theta_{n}}\end{array}\right)+\left(\begin{array}{c}
%H_{l}^{(2)}(k\rho_{n})\\
%\pm iH_{l+1}^{(2)}(k\rho_{n})e^{i\theta_{n}}\end{array}\right).\]

%\section*{\textcolor{black}{T-Matrix Scattering Theory for Graphene}}

Using the above results, we now derive the $t-$matrix for scattering of arbitrary incident waves of energy $E$ from a scatterer at $\vec{r}_{n}$.  The derivation follows that used for deriving the $t-$matrix in two-dimensional electron gases with Rashba spin$-$orbit coupling\cite{Walls06a}.
Writing the incident wave at the scatterer as $\Phi^{in}_{\pm\vec{K}}(\vec{r}_{n},\vec{k},\pm)=e^{i\vec{k}\cdot\vec{r}_{n}}|\theta_{\vec{k}},\pm\rangle_{\pm\vec{K}}$ and using the fact that $_{\pm\vec{K}}\langle\theta_{\vec{k}},\alpha|\theta_{\vec{k}},\beta\rangle_{\pm\vec{K}}=\delta_{\alpha\beta}$ and $\hat{L}_{\text{sgn}(l)}^{|l|}e^{i\vec{k}\cdot\vec{r}}=e^{il\theta_{\vec{k}}}e^{i\vec{k}\cdot\vec{r}}$
where sgn$(l)$ gives the sign of $l$, Eq.~(\ref{eq:scatteredwave}) can be rewritten as:
\begin{eqnarray}
\Psi^{s}_{\pm\vec{K}}(\vec{r},\vec{k},\pm)&=&\sqrt{k}\sum_{l=-\infty}^{\infty}s_{l}\left[e^{il\theta_{n}}\mathcal{H}_{l,\pm\vec{K}}^{(1)}(\vec{\rho}_{n},\vec{k},\pm)_{\pm\vec{K}}\langle\theta_{\vec{k}},\pm|\right]|\theta_{\vec{k}},\pm\rangle_{\pm\vec{K}}\left[i^{l}e^{-il\theta_{\vec{k}}}e^{i\vec{k}\cdot\vec{r}_{n}}\right]\nonumber\\
&=&\sum_{l=-\infty}^{\infty}\frac{i^{l}s_{l}e^{il\theta_{n}}}{2}\left(\begin{array}{cc}H^{(1)}_{l}(k\rho_{n})&\pm H^{(l)}_{l}(k\rho_{n})e^{-i\theta_{\vec{k}}}\\
\pm i H_{l+1}^{(1)}(k\rho_{n})e^{i\theta_{n}}&i H_{l+1}^{(1)}(k\rho_{n})e^{i(\theta_{n}-\theta_{\vec{k}})}\end{array}\right)_{\pm\vec{K}}|\theta_{\vec{k}},\pm\rangle_{\pm\vec{K}}\left[i^{l}e^{-il\theta_{\vec{k}}}e^{i\vec{k}\cdot\vec{r}_{n}}\right]\nonumber\\
&=&\sum_{l=-\infty}^{\infty}\frac{i^{l}s_{l}e^{il\theta_{n}}}{2}\left(\begin{array}{cc}H^{(1)}_{l}(k\rho_{n})&\pm H^{(l)}_{l}(k\rho_{n})\hat{L}_{-}\\
\pm i H^{(1)}_{l+1}(k\rho_{n})e^{i\theta_{n}}& iH_{l+1}^{(1)}(k\rho_{n})e^{i\theta_{n}}\hat{L}_{-}\end{array}\right)_{\pm\vec{K}}\hat{L}^{|l|}_{\text{sgn}(-l)}\Phi^{in}_{\pm\vec{K}}(\vec{r}_{n},\vec{k},\pm)\nonumber\\
&=&\sum_{l=-\infty}^{\infty}\frac{i^{l}e^{il\theta_{n}}}{2}\left(\begin{array}{cc}
H^{(1)}_{l}(k\rho_{n})&\mp i H^{(1)}_{l-1}(k\rho_{n})e^{-i\theta_{n}}\\
\pm i H_{l+1}^{(1)}(k\rho_{n})e^{i\theta_n}&H_{l}^{(1)}(k\rho_{n})\end{array}\right)_{\pm\vec{K}}\left(\begin{array}{cc}s_{l}&0\\
0&s_{l-1}\end{array}\right)_{\pm\vec{K}}\hat{L}^{|l|}_{\text{sgn}(-l)}\Phi^{in}_{\pm\vec{K}}(\vec{r}_{n},\vec{k},\pm)\label{eq:next2}
\end{eqnarray}
Eq.~(\ref{eq:next2}) can be regrouped into terms with the same scattering amplitude, $s_{l}$ [note from Eq.~(\ref{eq:phaseshift}), $s_{l}=s_{-(l+1)}$].  Using the fact that any wavefunction, $\Psi_{\pm\vec{K}}$, that is a solution to Eq.~(\ref{eq:h}) satisfies the following relations:
\begin{eqnarray}
\left(\begin{array}{cc}1&\mp\hat{L}_{-}\\\mp\hat{L}_{+}&1\end{array}\right)_{\pm\vec{K}}\Psi_{\pm\vec{K}}=\left(\begin{array}{cc}\mp \hat{L}_{-}^{n}&\hat{L}_{-}^{n+1}\\ \mp\hat{L}_{-}^{n}&\hat{L}^{n+1}_{-}\end{array}\right)_{\pm\vec{K}}\Psi_{\pm\vec{K}}=\left(\begin{array}{cc}\hat{L}^{n+1}_{+}&\mp \hat{L}_{+}^{n}\\ \hat{L}^{n+1}_{+}&\mp\hat{L}_{+}^{n}\end{array}\right)_{\pm\vec{K}}\Psi_{\pm\vec{K}}=0
\label{eq:graphrel}
\end{eqnarray}
Eq.~(\ref{eq:next2}) can be rewritten as:
\begin{eqnarray}
\Psi^{s}_{\pm\vec{K}}(\vec{r},\vec{k},\pm)&=&\sum_{l=0}^{\infty}\frac{4i\hbar\nu_{F}s_{l}}{k}\widehat{G}_{l,\pm\vec{K}}(\vec{r},\vec{r}_{n},E)\widehat{T}_{l,\pm\vec{K}}\left[\Phi^{in}_{\pm\vec{K}}(\vec{r}_{n},\vec{k},\pm)\right]
\end{eqnarray}
%\left[\frac{4is_{0}\hbar\nu_{F}}{k}\widehat{G}^{\text{clean}}_{0,\pm\vec{K}}(\vec{r},\vec{r}_{n},E)+\sum_{l=1}^{\infty}\frac{4is_{l}\hbar\nu_{F}}{k}\left(\widehat{G}^{\text{clean},-}_{-l,\pm\vec{K}}(\vec{r},\vec{r}_{n},E)\hat{L}^{l}_{+}+\widehat{G}^{\text{clean},+}_{+l,\pm\vec{K}}(\vec{r},\vec{r}_{n},E)\hat{L}^{l}_{-}\right)\right]\Phi^{in}_{\pm\vec{K}}(\vec{r}_{n},\vec{k},\pm)
%\end{eqnarray}
where
\begin{eqnarray}
\widehat{G}_{l,\pm\vec{K}}(\vec{r},\vec{r}_{n},E)&=&-\frac{i^{l+1}k}{4\hbar\nu_{F}}\left(\begin{array}{cc}H^{(1)}_{l}(k\rho_{n})e^{il\theta_{n}}&\pm i H^{(1)}_{l+1}(k\rho_{n})e^{-i(l+1)\theta_{n}}\\
 \pm i H_{l+1}^{(1)}(k\rho_{n})e^{i(l+1)\theta_{n}}&H_{l}^{(1)}(k\rho_{n})e^{-il\theta_{n}}\end{array}\right)_{\pm\vec{K}}\nonumber\\
 &=&-\frac{ik}{4\hbar\nu_{F}}\left(\begin{array}{cc}\hat{L}^{l}_{+}[H_{0}^{(1)}(k\rho_{n})]&\mp i\hat{L}^{l}_{-}[H^{(1)}_{-1}(k\rho_{n})e^{-i\theta_{n}}]\\ \pm i\hat{L}^{l}_{+}[H_{1}^{(1)}(k\rho_{n})e^{i\theta_{n}}]&\hat{L}^{l}_{-}[H_{0}^{(1)}(k\rho_{n})]\end{array}\right)_{\pm\vec{K}}
 \label{eq:greensing}
 \end{eqnarray}
and $\widehat{T}_{l,\pm\vec{K}}$ is the $l$-partial wave $t-$matrix operator given by:
\begin{eqnarray}
\widehat{T}_{l,\pm\vec{K}}=\left(\begin{array}{cc}\hat{L}_{-}^{l}&0\\0&\hat{L}_{+}^{l}\end{array}\right)_{\pm\vec{K}}
\end{eqnarray}
Note that $\widehat{G}_{0,\pm{K}}(\vec{r},\vec{r}_{n},E)$ is simply the Green's function for single layer graphene about the $\pm\vec{K}$ Dirac point\cite{Braun08}.

In a similar manner, the total Green's function, $\widehat{G}_{\pm\vec{K}}(\vec{r},\vec{r}',E)$, can be calculated and is given as [using the above notation]:
\begin{eqnarray}
\widehat{G}_{\pm\vec{K}}(\vec{r},\vec{r}',E)&=&\left(\begin{array}{cc}G_{11}(\vec{r},\vec{r}',E)&G_{12}(\vec{r},\vec{r}',E)\\ G_{21}(\vec{r},\vec{r}',E)&G_{22}(\vec{r},\vec{r}',E)\end{array}\right)_{\pm\vec{K}}\nonumber\\
&=&\widehat{G}_{0,\pm\vec{K}}(\vec{r},\vec{r}',E)+\sum_{l=0}^{\infty}\frac{4i\hbar\nu_{F}s_{l}}{k}\widehat{G}_{l,\pm\vec{K}}(\vec{r},\vec{r}_{n},E)\widehat{T}_{l,\pm\vec{K}}\left[\widehat{G}_{\pm\vec{K}}(\vec{r}_{n},\vec{r}',E)\right]\nonumber\\
&=&\widehat{G}_{0,\pm\vec{K}}(\vec{r},\vec{r}',E)+\sum_{l=0}^{\infty}\frac{4is_{l}\hbar\nu_{F}}{k}\widehat{G}_{l,\pm\vec{K}}(\vec{r},\vec{r}_{n},E)\left(\begin{array}{cc}\hat{L}_{-}^{l}\left[G_{11}(\vec{r}_{n},\vec{r}',E)\right]&\hat{L}_{-}^{l}\left[G_{12}(\vec{r}_{n},\vec{r}',E)\right]\\ \hat{L}_{+}^{l}\left[G_{21}(\vec{r}_{n},\vec{r}',E)\right]&\hat{L}_{+}^{l}\left[G_{22}(\vec{r}_{n},\vec{r}',E)\right]\end{array}\right)_{\pm\vec{K}}\nonumber\\
\label{eq:greensosing}
\end{eqnarray}
For a single scatterer at $\vec{r}_{n}$,
\begin{eqnarray}
\left(\begin{array}{cc}\hat{L}_{-}^{l}\left[G_{11}(\vec{r}_{n},\vec{r}',E)\right]&\hat{L}_{-}^{l}\left[G_{12}(\vec{r}_{n},\vec{r}',E)\right]\\ \hat{L}_{+}^{l}\left[G_{21}(\vec{r}_{n},\vec{r}',E)\right]&\hat{L}_{+}^{l}\left[G_{22}(\vec{r}_{n},\vec{r}',E)\right]\end{array}\right)_{\pm\vec{K}}&=&\widehat{T}_{l,\pm\vec{K}}\left[\widehat{G}_{0,\pm\vec{K}}(\vec{r}_{n},\vec{r}',E)\right]\nonumber\\
&=&\frac{(-i)^{l+1}k}{4\hbar\nu_{F}}
\left(\begin{array}{cc}H^{(1)}_{l}(k\rho'_{n})e^{-il\theta'_{n}}&\mp i H^{(1)}_{l+1}(k\rho'_{n})e^{-i(l+1)\theta'_{n}}\\
 \mp i H_{l+1}^{(1)}(k\rho'_{n})e^{i(l+1)\theta'_{n}}&H_{l}^{(1)}(k\rho'_{n})e^{il\theta'_{n}}\end{array}\right)_{\pm\vec{K}}
 \label{eq:greenv}
\end{eqnarray}
where $\rho'_{n}=|\vec{r}_{n}-\vec{r}'|$ and $e^{\pm il\theta'_{n}}=\left(\frac{(\vec{r}'-\vec{r}_{n})\cdot(\hat{x}\pm i\hat{y})}{\rho'_{n}}\right)^{l}$.  In Eq.~(\ref{eq:greenv}), we used the relation that $\hat{L}^{l'}_{\pm}\left[H^{(1)}_{l}(kr)e^{il\theta}\right]=i^{\pm l}H^{(1)}_{l\pm l'}(kr)e^{i(l\pm l')\theta}$.  Inserting Eq.~ (\ref{eq:greenv}) into Eq.~(\ref{eq:greensosing}) completely determines the Green's function in the presence of a single scatterer.

The LDOS, $\rho_{\pm\vec{K}}(\vec{r},E)$, which is an important quantity that has been previously measured in STM experiments on graphene\cite{Rutter07,Brihuega08}, can be calculated from the Green's function using the relation\cite{Tersoff85} $\rho_{\pm\vec{K}}(\vec{r},E)=-\frac{1}{\pi}\text{Im}\left[\widehat{G}_{\pm\vec{K}}(\vec{r},\vec{r},E)\right]$.
With Eq.~(\ref{eq:greensosing}), $\rho_{\pm\vec{K}}(\vec{r},E)$ for single layer graphene in the presence of a scatterer at $\vec{r}_{n}$ is given by:
\begin{eqnarray}
\rho_{\pm\vec{K}}(\vec{r},E)&=&\frac{k}{4\pi\hbar\nu_{F}}\left(1+\sum_{l=0}^{\infty}\text{Im}\left[is_{l}\left(\left(H_{l}^{(1)}(k\rho_{n})\right)^{2}+\left(H_{l+1}^{(1)}(k\rho_{n})\right)^{2}\right)\right]\right)\left(\begin{array}{cc}1&0\\0&1\end{array}\right)_{\pm\vec{K}}\nonumber\\
&=&\frac{k}{4\pi\hbar\nu_{F}}\left(1+\sum_{l=0}^{\infty}\text{Im}\left[is_{l}\left(\left(H_{l}^{(1)}(k\rho_{n})\right)^{2}+\left(H_{l+1}^{(1)}(k\rho_{n})\right)^{2}\right)\right]\right)\left(\left|\Phi^{\text{lat}}_{A}(\vec{r},\pm\vec{K})\right|^{2}+\left|\Phi^{\text{lat}}_{B}(\vec{r},\pm\vec{K})\right|^{2}\right)\nonumber\\
&=&\rho^{\text{clean}}_{\pm\vec{K}}(\vec{r},E)+\delta\rho_{\pm\vec{K}}(\vec{r},E)\nonumber\\
&=&\rho^{\text{clean}}_{\pm\vec{K}}(\vec{r},E)\left(1+\frac{\delta\rho_{\pm\vec{K}}(\vec{r},E)}{\rho^{\text{clean}}_{\pm\vec{K}}(\vec{r},E)}\right)
\label{eq:singscatt2}
\end{eqnarray}
where $\Phi^{\text{lat}}_{A}(\vec{r},\pm\vec{K})$ and $\Phi^{\text{lat}}_{B}(\vec{r},\pm\vec{K})$ are the lattice wave functions [Eq.~ (\ref{eq:latticewave})] evaluated at $\vec{r}$, \\ $\rho^{\text{clean}}_{\pm\vec{K}}(\vec{r},E)=\frac{k}{4\pi\hbar\nu_{F}}\left(\left|\Phi^{\text{lat}}_{A}(\vec{r},\pm\vec{K})\right|^{2}+\left|\Phi^{\text{lat}}_{B}(\vec{r},\pm\vec{K})\right|^{2}\right)$ is the LDOS for single layer graphene about the $\pm\vec{K}$ Dirac points in the absence of any scatterers, $\delta\rho_{\pm\vec{K}}(\vec{r},E)$ is the change in the LDOS due to the scatterer, and\begin{eqnarray} \frac{\delta\rho_{\pm\vec{K}}(\vec{r},E)}{\rho^{\text{clean}}_{\pm\vec{K}}(\vec{r},E)}
=\sum_{l=0}^{\infty}\text{Im}\left[is_{l}\left(\left(H_{l}^{(1)}(k\rho_{n})\right)^{2}+\left(H_{l+1}^{(1)}(k\rho_{n})\right)^{2}\right)\right]
\label{eq:env}\end{eqnarray} is the ``envelope" function of the Friedel oscillations in the LDOS.

For $k\rho_{n}\gg 1$, $\left(H^{(1)}_{l}(k\rho_{n})\right)^{2}\approx \frac{2}{\pi k\rho_{n}}e^{2ik\rho_{n}-i\frac{\pi}{2}-il\pi}+O\left[\left(\frac{1}{k\rho_{n}}\right)^{2}\right]$ and so in this limit, Eq.~(\ref{eq:env}) becomes:
\begin{eqnarray}
\frac{\delta\rho_{\pm\vec{K}}(\vec{r},E)}{\rho^{\text{clean}}_{\pm\vec{K}}(\vec{r},\vec{r},E)}\approx\frac{2}{\pi k\rho_{n}}\sum_{l=0}^{\infty}\text{Im}\left[s_{l}(-1)^{l}\left(e^{i2k\rho_{n}}-e^{i2k\rho_{n}}+O\left[\frac{1}{k\rho_{n}}\right]\right)\right] = O\left[\left(\frac{1}{k\rho_{n}}\right)^{2}\right]
\label{eq:decayF}
\end{eqnarray}

\begin{figure}
\begin{centering}
\includegraphics[scale=0.5]{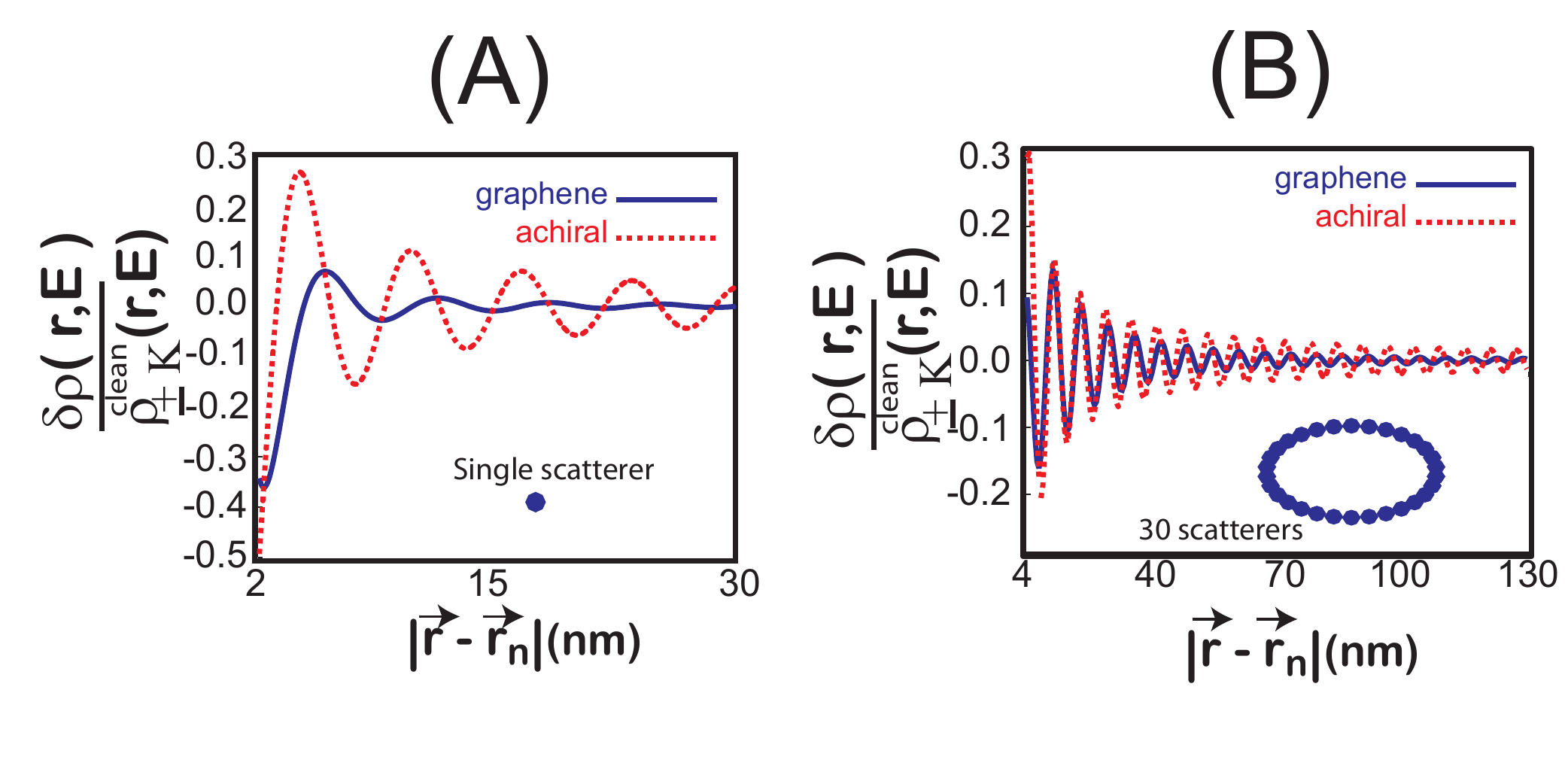}\caption{[Color Online] The calculated relative change in the LDOS, $\frac{\delta\rho(\vec{r},E)}{\rho^{\text{clean}}_{\pm\vec{K}}(\vec{r},E)}$ in the presence of (A) a single scatterer at $\vec{r}_{n}$ and (B) an elliptical arrangement of $N=30$ identical scatterers [major and minor axes of 20 nm and 10 nm respectively] in both graphene [solid, blue curve] and in a 2D electron gas (2DEG) or achiral system [dotted, red curve] as a function of (A) the distance from the single scatter  or as a function of (B) the distance from the outermost scatterer along the semimajor axis of the ellipse.  In both calculations, $a=10$~\AA, $V_{0}=4$ eV, the scattering amplitudes were given by $s_{l}$ in Eq. (\ref{eq:phaseshift}), and $k=0.485~\text{nm}^{-1}$ [corresponding to $E=0.32$ eV for graphene].  In both (A) and (B), the first five partial waves [$l=0$ to $l=4$] were included in the calculations, which provides an accuracy of the numerical results to within 1 percent. (A) For the single scatterer case, the Friedel oscillations in the LDOS decrease as $(k\rho_{n})^{-2}$ for the graphene/chiral case [blue, solid curve], whereas they decrease as $(k\rho_{n})^{-1}$ in a 2DEG/achiral case [red, dotted curve].  (B)  Due to multiple scattering within the elliptical array of scatterers, the Friedel oscillations in the LDOS decrease at roughly the same rate as those found in a 2DEG and can be observed at distances up to 100 nm from the scatterers.        \label{fig:figure2}}
\par\end{centering}

\end{figure}

Thus the Friedel oscillations in $\delta\rho_{\pm\vec{K}}$ decay as $(k\rho_{n})^{-2}$ in graphene\cite{Bena08}, whereas for $k\rho_{n}\gg 1$,  the envelope of the Friedel oscillations in 2D electron gas (2DEG) or achiral systems\cite{Mariani07,Cheianov07,Bena08} decay as $\sum_{l=0}^{\infty}\text{Im}\left[is_{l}\left(H_{l}^{(1)}(k\rho_{n})\right)^{2}\right]\rightarrow\frac{2}{\pi k\rho_{n}}\sum_{l=0}^{\infty}\text{Im}\left[s_{l}(-1)^{l}e^{2ik\rho_{n}}\right]$. This feature is illustrated in the numerical simulation of $\frac{\delta\rho(\vec{r},E)}{\rho^{\text{clean}}_{\pm\vec{K}}(\vec{r},E)}$ shown in Fig.~\ref{fig:figure2}(A).  From Eq.~ (\ref{eq:decayF}), the $(k\rho_{n})^{-2}$ dependence in the Friedel oscillations is due to the destructive interference between $H_{l}^{(1)}(k\rho_{n})$ and $H^{(1)}_{l+1}(k\rho_{n})$ along a collinear scattering trajectory from the position $\vec{r}$, to the scatterer and then back $[\vec{r}\rightarrow\vec{r}_{n}\rightarrow\vec{r}]$.  While it is tempting to attribute the $(k\rho_{n})^{-2}$ dependence of the Friedel oscillations in graphene to the absence\cite{Ando98} of intravalley backscattering ($\vec{k}\nrightarrow -\vec{k}$), this can not be the complete explanation behind the decay found in Eq.~(\ref{eq:decayF}).  As a case in point, the calculated LDOS [using the multiple scattering theory developed in the next section] outside an elliptical array of $N=30$ scatterers [semimajor and semiminor axis of 20 nm and 10 nm respectively] and along the direction of the semimajor axis is shown in Fig.~\ref{fig:figure2}(B).  Although backscattering [$\vec{k}\rightarrow -\vec{k}$] from an elliptical array of scatterers is still prohibited in graphene by time-reversal symmetry\cite{Ando98}, the Friedel oscillations for graphene extend out a hundred nanometers from the scatterer unlike that found for a single scatterer [Fig.~\ref{fig:figure2}(A)] where the Friedel oscillations would have died out within tens of nanometers from the scatterer array.  For the elliptical array of scatterers, there exist noncollinear multiple scattering trajectories that prevent the complete destructive interference between the $H_{l}^{(1)}(k\rho_{n})$ and $H^{(1)}_{l+1}(k\rho_{n})$ components of the $l^{th}$-partial wave, thereby leading to a slower decay of the Friedel oscillations.

\section{Multiple scattering in graphene}
The above theory can be extended to investigate multiple scattering from $N$ scatterers in graphene.  However, for multiple scatterers, it is computationally unfeasible to consider all partial waves.  Fortunately for most physical potentials, the scattering amplitude, $s_{l}$, is nonneglible for only a small subset of $l$, thereby justifying the use of only a few partial waves in the scattering calculations.  In the following theory, only the first $l_{max}+1$ partial waves [$l=0$ to $l=l_{max}$] will be considered [a discussion of the proper choice of $l_{max}$ will be provided later in the paper].   Denoting the position of the $j^{th}$ scatterer by $\vec{r}_{j}$ and the $l^{th}$ partial wave scattering amplitude from scatterer $j$ as $s_{l}^{(j)}$,
the total Green's function at energy $E=\hbar\nu_{F}k\geq 0$, can be written as [$\vec{r},\vec{r}'\neq\vec{r}_{j}$ for $j=1$ to $j=N$]:
\begin{eqnarray}
\widehat{G}_{\pm\vec{K}}(\vec{r},\vec{r}',E)&=&\widehat{G}_{0,\pm\vec{K}}(\vec{r},\vec{r}',E)+\sum_{j=1}^{N}\sum_{l=0}^{l_{max}}\frac{4i\hbar\nu_{F}s_{l}^{(j)}}{k}\widehat{G}_{l,\pm\vec{K}}(\vec{r},\vec{r}_{j},E)\widehat{T}_{l,\pm\vec{K}}\left[\widehat{G}_{\pm\vec{K}}(\vec{r}_{j},\vec{r}',E)\right]
\label{eq:greenprelim}
\end{eqnarray}
In Eq.~(\ref{eq:greenprelim}), knowledge of $\widehat{T}_{l,\pm\vec{K}}\left[\widehat{G}_{\pm\vec{K}}(\vec{r}_{j},\vec{r}',E)\right]$ for each partial wave $l=0$ to $l=l_{max}$ and for each scatterer $j=1$ to $j=N$ completely determines the total Green's function, $\widehat{G}_{\pm\vec{K}}(\vec{r},\vec{r}',E)$.  In this case, there are $4N(l_{max}+1)$ unknowns that can be determined self-consistently from the corresponding Foldy$-$Lax\cite{Foldy45,Lax51} equations derived from Eq.~(\ref{eq:greenprelim}):
\begin{eqnarray}
\widehat{T}_{l',\pm\vec{K}}\left[\widehat{G}_{\pm\vec{K}}(\vec{r}_{n},\vec{r}',E)\right]&=&\widehat{T}_{l',\pm\vec{K}}\left[\widehat{G}_{0,\pm\vec{K}}(\vec{r}_{n},\vec{r}',E)\right]+\sum_{j\neq n}\sum_{l=0}^{l_{max}}\frac{4i\hbar\nu_{F}s^{(j)}_{l}}{k}\widehat{T}_{l',\pm\vec{K}}\left[\widehat{G}_{l,\pm\vec{K}}(\vec{r}_{n},\vec{r}_{j},E)\right]\widehat{T}_{l,\pm\vec{K}}
\left[\widehat{G}_{\pm\vec{K}}(\vec{r}_{j},\vec{r}',E)\right]
\label{eq:selfgreen}
\end{eqnarray}
for $n=1$ to $n=N$ and $l'=0$ to $l'=l_{max}$.
where
\begin{eqnarray}
T_{l',\pm\vec{K}}\left[\widehat{G}_{l,\pm\vec{K}}(\vec{r}_{n},\vec{r}_{j})\right]&=&-\frac{ik}{4\hbar\nu_{F}}i^{l+l'}\left(\begin{array}{cc}(-1)^{l'}H^{(1)}_{l-l'}(kr_{nj})e^{i(l-l')\theta_{nj}}   &\pm iH^{(1)}_{l+l'+1}(k\rho_{nj})e^{-i\theta_{nj}(l+l'+1)}\\
\pm iH^{(1)}_{l+l'+1}(kr_{nj})e^{i\theta_{nj}(l+l'+1)}&(-1)^{l}H^{(1)}_{l'-l}(kr_{nj})e^{i\theta_{nj}(l'-l)}\end{array}\right)_{\pm\vec{K}}
\end{eqnarray}

Defining the following $2(l_{max}+1)\times 2$ matrix for each scatterer $j$:
\begin{eqnarray}
\widehat{\widehat{{T}}}\widehat{\widehat{G}}^{(j)}(\vec{r}_{j},\vec{r}')_{\pm\vec{K}}&=&\left(\begin{array}{cc}\widehat{T}_{0,\pm\vec{K}}[\widehat{G}_{\pm\vec{K}}(\vec{r}_{j},\vec{r}',E)]\\
\widehat{T}_{1,\pm\vec{K}}[\widehat{G}_{\pm\vec{K}}(\vec{r}_{j},\vec{r}',E)]\\
\widehat{T}_{2,\pm\vec{K}}[\widehat{G}_{\pm\vec{K}}(\vec{r}_{j},\vec{r}',E)]\\
\vdots\\
\widehat{T}_{l_{max},\pm\vec{K}}\left[\widehat{G}_{\pm\vec{K}}(\vec{r}_{j},\vec{r}',E)\right]
\end{array}\right),\widehat{\widehat{T}}\widehat{\widehat{G}}^{(j)}_{0}(\vec{r}_{j},\vec{r}')=\left(\begin{array}{cc}\widehat{T}_{0,\pm\vec{K}}[\widehat{G}_{0,\pm\vec{K}}(\vec{r}_{j},\vec{r}',E)]\\
\widehat{T}_{1,\pm\vec{K}}[\widehat{G}_{0,\pm\vec{K}}(\vec{r}_{j},\vec{r}',E)]\\
\widehat{T}_{2,\pm\vec{K}}[\widehat{G}_{0,\pm\vec{K}}(\vec{r}_{j},\vec{r}',E)]\\
\vdots\\
\widehat{T}_{l_{max},\pm\vec{K}}[\widehat{G}_{0,\pm\vec{K}}(\vec{r}_{j},\vec{r}',E)]
\end{array}\right)
\end{eqnarray}
and the following $2N(l_{max}+1)\times 2$ matrices:
\begin{eqnarray}
\mathbf{\widehat{\widehat{T}}}\mathbf{\widehat{\widehat{G}}}_{\pm\vec{K}}(\vec{r}')=\left(\begin{array}{c}
\widehat{\widehat{T}}\widehat{\widehat{G}}^{(1)}_{\pm\vec{K}}(\vec{r}_{1},\vec{r}')\\
\widehat{\widehat{T}}\widehat{\widehat{G}}^{(2)}_{\pm\vec{K}}(\vec{r}_{2},\vec{r}')\\
\vdots\\
\widehat{\widehat{T}}\widehat{\widehat{G}}^{(N)}_{\pm\vec{K}}(\vec{r}_{N},\vec{r}')
\end{array}\right),
\mathbf{\widehat{\widehat{T}}}\mathbf{\widehat{\widehat{G}}}_{0,\pm\vec{K}}(\vec{r}')=\left(\begin{array}{c}
\widehat{\widehat{T}}\widehat{\widehat{G}}^{(1)}_{0,\pm\vec{K}}(\vec{r}_{1},\vec{r}')\\
\widehat{\widehat{T}}\widehat{\widehat{G}}^{(2)}_{0,\pm\vec{K}}(\vec{r}_{2},\vec{r}')\\
\vdots\\
\widehat{\widehat{T}}\widehat{\widehat{G}}^{(N)}_{0,\pm\vec{K}}(\vec{r}_{N},\vec{r}')
\end{array}\right)
\end{eqnarray}
Eqs. (\ref{eq:selfgreen}) can be written compactly for $n=1$ to $n=N$ and for $l'=0$ to $l'=l_{max}$ as:
\begin{eqnarray}
\mathbf{\widehat{\widehat{T}}}\mathbf{\widehat{\widehat{G}}}_{\pm\vec{K}}(\vec{r}')&=&\left(\mathbf{\widehat{\widehat{1}}}-\mathbf{\widehat{\widehat{TT}}}\right)^{-1}\mathbf{\widehat{\widehat{T}}}\mathbf{\widehat{\widehat{G}}}_{0,\pm\vec{K}}(\vec{r}')
\label{eq:grenbo}
\end{eqnarray}
where
$\mathbf{\widehat{\widehat{1}}}$ is the $2N(l_{max}+1)\times 2N(l_{max}+1)$ identity matrix and $\mathbf{\widehat{\widehat{TT}}}$ is a $2N(l_{max}+1)\times 2N(l_{max}+1)$ matrix given by:
\begin{eqnarray}
\mathbf{\widehat{\widehat{TT}}}&=&\left(\begin{array}{ccccc}\mathbf{\widehat{0}}&\widehat{\widehat{TT}}_{\pm\vec{K}}(\vec{r}_{1},\vec{r}_{2})&
\widehat{\widehat{TT}}_{\pm\vec{K}}(\vec{r}_{1},\vec{r}_{3})&\hdots&\widehat{\widehat{TT}}_{\pm\vec{K}}(\vec{r}_{1},\vec{r}_{N})\\
\widehat{\widehat{TT}}_{\pm\vec{K}}(\vec{r}_{2},\vec{r}_{1})&\mathbf{\widehat{0}}&\widehat{\widehat{TT}}_{\pm\vec{K}}(\vec{r}_{2},\vec{r}_{3})&\hdots&\widehat{\widehat{TT}}_{\pm\vec{K}}(\vec{r}_{2},\vec{r}_{N})\\
\vdots&\vdots&\vdots&\ddots&\vdots\\
\widehat{\widehat{TT}}_{\pm\vec{K}}(\vec{r}_{N},\vec{r}_{1})&\widehat{\widehat{TT}}_{\pm\vec{K}}(\vec{r}_{N},\vec{r}_{2})&\widehat{\widehat{TT}}_{\pm\vec{K}}(\vec{r}_{N},\vec{r}_{3})&\hdots&\mathbf{\widehat{0}}
\end{array}
\right)
\end{eqnarray}
where \begin{eqnarray}
\widehat{\widehat{TT}}_{\pm\vec{K}}(\vec{r}_{n},\vec{r}_{j})&=&\left(\begin{array}{ccccc}\widehat{G}_{0,\pm\vec{K}}(\vec{r}_{n},\vec{r}_{j})
&\widehat{G}_{1,\pm\vec{K}}(\vec{r}_{n},\vec{r}_{j})
&\widehat{G}_{2,\pm\vec{K}}(\vec{r}_{n},\vec{r}_{j})
&\hdots
&\widehat{G}_{l_{max},\pm\vec{K}}(\vec{r}_{n},\vec{r}_{j})\\
\widehat{T}_{1,\pm\vec{K}}\left[\widehat{G}_{0,\pm\vec{K}}(\vec{r}_{n},\vec{r}_{j})\right]
&\widehat{T}_{1,\pm\vec{K}}\left[\widehat{G}_{1,\pm\vec{K}}(\vec{r}_{n},\vec{r}_{j})\right]
&\widehat{T}_{1,\pm\vec{K}}\left[\widehat{G}_{2,\pm\vec{K}}(\vec{r}_{n},\vec{r}_{j})\right]
&\hdots
&\widehat{T}_{1,\pm\vec{K}}\left[\widehat{G}_{l_{max},\pm\vec{K}}(\vec{r}_{n},\vec{r}_{j})\right]\\
\widehat{T}_{2,\pm\vec{K}}\left[\widehat{G}_{0,\pm\vec{K}}(\vec{r}_{n},\vec{r}_{j})\right]
&\widehat{T}_{2,\pm\vec{K}}\left[\widehat{G}_{1,\pm\vec{K}}(\vec{r}_{n},\vec{r}_{j})\right]
&\widehat{T}_{2,\pm\vec{K}}\left[\widehat{G}_{2,\pm\vec{K}}(\vec{r}_{n},\vec{r}_{j})\right]
&\hdots
&\widehat{T}_{2,\pm\vec{K}}\left[\widehat{G}_{l_{max},\pm\vec{K}}(\vec{r}_{n},\vec{r}_{j})\right]\\
\vdots&\vdots&\vdots&\ddots&\vdots\\
\widehat{T}_{l_{max},\pm\vec{K}}\left[\widehat{G}_{0,\pm\vec{K}}(\vec{r}_{n},\vec{r}_{j})\right]
&\widehat{T}_{l_{max},\pm\vec{K}}\left[\widehat{G}_{1,\pm\vec{K}}(\vec{r}_{n},\vec{r}_{j})\right]
&\widehat{T}_{l_{max},\pm\vec{K}}\left[\widehat{G}_{2,\pm\vec{K}}(\vec{r}_{n},\vec{r}_{j})\right]
&\hdots
&\widehat{T}_{l_{max},\pm\vec{K}}\left[\widehat{G}_{l_{max},\pm\vec{K}}(\vec{r}_{n},\vec{r}_{j})\right]
\end{array}\right)\mathbf{\widehat{S}}_{l_{max}}^{(j)}\nonumber\\
\end{eqnarray}
and
\begin{eqnarray}
\mathbf{\widehat{S}}_{l_{max}}^{(j)}&=&\frac{4i\hbar\nu_{F}}{k}\left(\begin{array}{ccccc}s^{(j)}_{0}&0&0&\hdots&0\\
0&s^{(j)}_{1}&0&\hdots&0\\
0&0&s^{(j)}_{2}&\hdots&0\\
\vdots&\vdots&\vdots&\ddots&\vdots\\
0&0&0&\hdots&s^{(j)}_{l_{max}}\end{array}
\right)\otimes\left(\begin{array}{cc}1&0\\0&1\end{array}\right)
\end{eqnarray}
Finally, $\widehat{G}_{\pm\vec{K}}(\vec{r},\vec{r}',E)$ in Eq.~(\ref{eq:greenprelim}) can be written compactly as:
\begin{eqnarray}
\widehat{G}_{\pm\vec{K}}(\vec{r},\vec{r}',E)&=&\widehat{G}_{0,\pm\vec{K}}(\vec{r},\vec{r}',E)+\mathbf{\widehat{\widehat{GG}}}(\vec{r})\mathbf{\widehat{\widehat{T}}\widehat{\widehat{G}}}(\vec{r}')\nonumber\\
&=&\widehat{G}_{0,\pm\vec{K}}(\vec{r},\vec{r}',E)+\mathbf{\widehat{\widehat{GG}}}(\vec{r})\left(\mathbf{\widehat{\widehat{1}}}-\mathbf{\widehat{\widehat{TT}}}\right)^{-1}\mathbf{\widehat{\widehat{T}}\widehat{\widehat{G}}}_{0}(\vec{r}')
\label{eq:Gfin}
\end{eqnarray}
where $\mathbf{\widehat{\widehat{{GG}}}}(\vec{r})$ is a $2\times 2N(l_{max}+1)$ matrix, $\mathbf{\widehat{\widehat{GG}}}(\vec{r})=\left[\widehat{\widehat{GG}}(\vec{r},\vec{r}_{1})\,\,\widehat{\widehat{GG}}(\vec{r},\vec{r}_{2})\,\,\hdots\widehat{\widehat{GG}}(\vec{r},\vec{r}_{N})\right]$
where
\begin{eqnarray}
\widehat{\widehat{GG}}(\vec{r},\vec{r}_{j})&=&\frac{4i\hbar\nu_{F}}{k}\left[\begin{array}{cccc}s^{(j)}_{0}\widehat{G}_{0,\pm\vec{K}}(\vec{r},\vec{r}_{j})
&s^{(j)}_{1}\widehat{G}_{1,\pm\vec{K}}(\vec{r},\vec{r}_{j})&\hdots&s^{(j)}_{l_{max}}\widehat{G}_{l_{max},\pm\vec{K}}(\vec{r},\vec{r}_{j})\end{array}\right]
\end{eqnarray}
Using $\widehat{G}_{\pm\vec{K}}(\vec{r},\vec{r}',E)$ in Eq.~(\ref{eq:Gfin}), the LDOS, $\rho_{\pm\vec{K}}(\vec{r},E)$, and all transport properties can be calculated in graphene in the presence of $N$ scatterers.
\section{Results and Discussion}
In this section, we use Eq.~(\ref{eq:Gfin}) to calculate the effects of quasiparticle interference on the LDOS in single layer graphene in the presence of multiple scatterers.  As the seminal work of the Eigler group\cite{Crommie93} demonstrated, quantum corrals comprised of adatoms placed atop a metal surface lead to quasiparticle interference, where changes in the LDOS mimic the behavior of an electron confined in a billiard with leaky walls\cite{Heller94,Fiete03,Barr10}.  While the quasiparticles in graphene cannot be confined by such corrals due to their Dirac-like nature (this phenomenon is a consequence of the Klein paradox\cite{Katsnelson06}), non-collinear multiple scattering trajectories in graphene can result in dramatic changes in the LDOS compared with single scattering trajectories, which was previously illustrated in Fig.~\ref{fig:figure2}.

\begin{figure}
\begin{centering}
\includegraphics[scale=0.35]{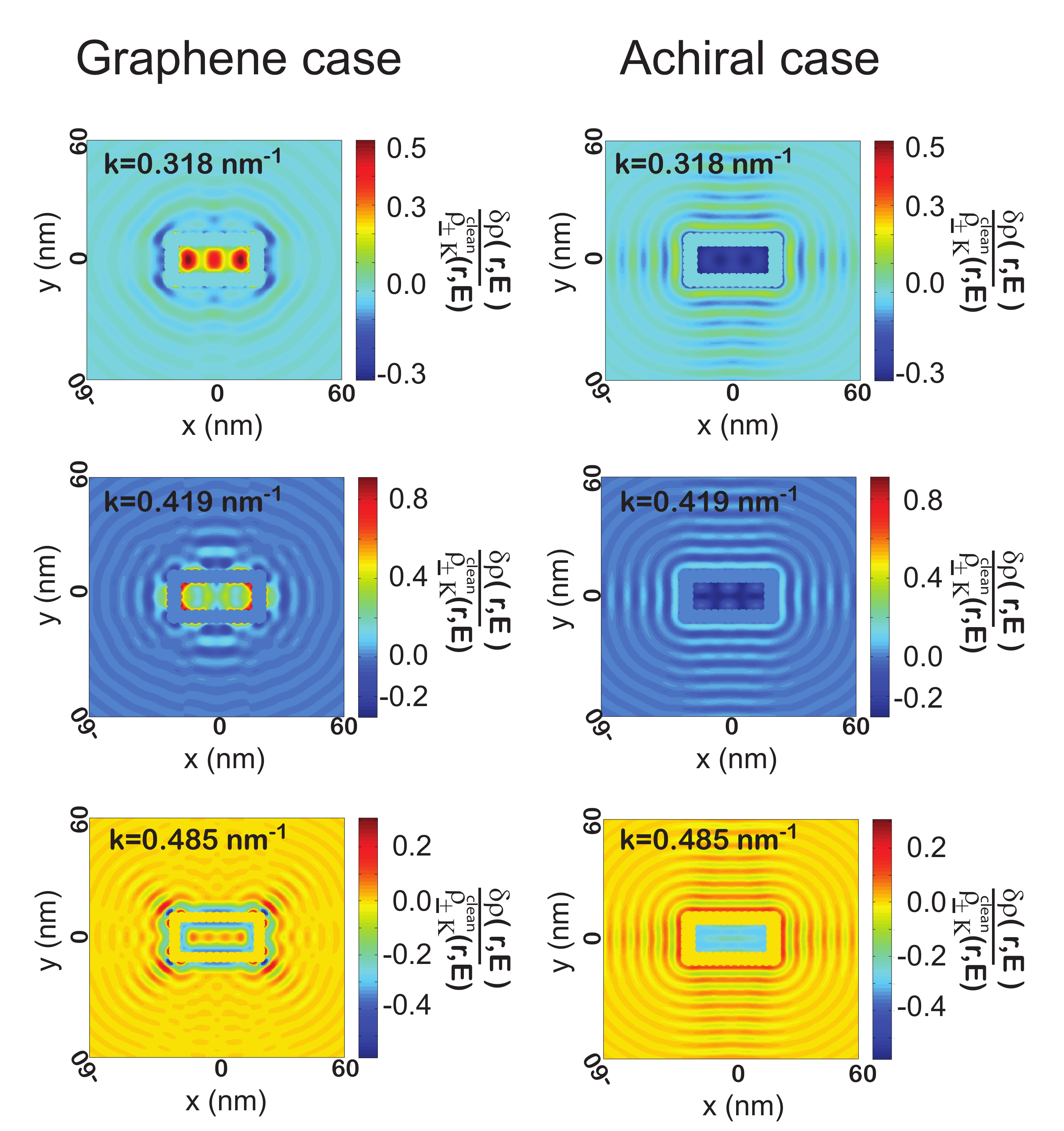}\caption{[Color Online] Theoretical calculations of $\frac{\delta\rho(\vec{r},E)}{\rho^{\text{clean}}_{\pm\vec{K}}(\vec{r},E)}$ for a 200~\AA~ by 400~\AA~ rectangular array of $N=36$ identical scatterers [$V=4$ eV, $a=10$~\AA] for wave vectors $k=0.318~\text{nm}^{-1}$ [top, $E=0.21$ eV for graphene], $k=0.419~\text{nm}^{-1}$ [middle, $E=0.2765$ eV for graphene], and $k=0.485~\text{nm}^{-1}$ [bottom, $E=0.32$ eV for graphene] for both a 2DEG [right, Eq.~(\ref{eq:Gfinac})] and graphene [left, Eq.~(\ref{eq:Gfin})].  In all simulations, $\frac{\delta\rho(\vec{r},E)}{\rho^{\text{clean}}_{\pm\vec{K}}(\vec{r},E)}$ was set to zero within 35~\AA~ of each scatterer, $s_{l}$ was calculated using Eq. (\ref{eq:phaseshift}) in both graphene and the 2DEG, and $l_{max}=4$ was used in all calculations in order to ensure a relative accuracy of calculations to within one percent.   Dramatic differences in the quasiparticle interference patterns between the graphene [left] and 2DEG [right] cases are observed due to the interplay between multiple scattering and the resulting pseudospin rotations         \label{fig:rect}}
\par\end{centering}
\end{figure}

In Fig.~\ref{fig:rect}, the relative change in the LDOS, $\frac{\delta\rho(\vec{r},E)}{\rho^{\text{clean}}_{\pm\vec{K}}(\vec{r},E)}$, is shown for a rectangular array [10 nm by 20 nm] of $N=36$ identical scatterers [$a=1$ nm and $V=4$ eV] in both graphene [left, calculated using Eq.~(\ref{eq:Gfin})] and in a 2DEG [right, calculated from Eq.~(\ref{eq:Gfinac})].  For better comparison between the 2DEG and graphene cases, the same magnitude of wave vector, $k$, and scattering amplitudes, $s_{l}$ [Eq.~ (\ref{eq:phaseshift})] were used in evaluating the LDOS for both the graphene [Eq.~(\ref{eq:Gfin})] and the 2DEG [Eq.~ (\ref{eq:Gfinac})] cases.  Therefore, the differences between calculated LDOS for the graphene and 2DEG cases arise solely from differences in the ``pure" Green's function in the absence of scatterers, the chiral $\widehat{G}_{0,\pm \vec{K}}(\vec{r},\vec{r}',E)\neq \widehat{G}_{0,\pm,\vec{K}}(\vec{r}',\vec{r},E)$ in Eq. (\ref{eq:greensing}) for graphene  vs. the achiral $\widehat{G}_{0}(\vec{r},\vec{r}',E)=\widehat{G}_{0}(\vec{r}',\vec{r},E)=-i\frac{m}{2\hbar^{2}}H^{(1)}_{0}\left(\sqrt{\frac{2mE}{\hbar^{2}}}|\vec{r}-\vec{r}'|\right)$ for the 2DEG.  For accurate calculations of $\widehat{G}(\vec{r},\vec{r},E)$ and hence the LDOS at $\vec{r}$, all scattering trajectories beginning and ending at $\vec{r}$ must be taken into account when closely spaced scatterers are present\cite{Fiete03}, as is done in Eq. (\ref{eq:Gfin}) and Eq. (\ref{eq:Gfinac}).

From Fig.~\ref{fig:rect}, a few general observations can be made.  First, the Friedel oscillations outside of the scatterer array are in general weaker in graphene than those found in the 2DEG, which is due to the fact that collinear backscattering trajectories [Fig.~\ref{fig:figure2}(A)] contribute little to the LDOS away from the scatterers.  Second, the shape of the Friedel oscillations outside the rectangular array of scatterers differs significantly between graphene and the 2DEG.  For the 2DEG, the Friedel oscillations are rectangular in shape away from the array of scatterers.  In graphene, however, the shape of the Friedel oscillations is not rectangular and is highly anisotropic.  Treating $\Phi^{\text{lat}}_{A}(\vec{r},\pm\vec{K})$ and $\Phi^{\text{lat}}_{B}(\vec{r},\pm\vec{K})$ as a pseudospin pair, multiple scattering trajectories result in pseudospin rotations that alter the quasiparticle intereference and thereby modulate the LDOS since the interference between two different trajectories is maximal only when the pseudospins from each trajectory are pointed along the same direction.  A similar result was predicted for the LDOS of quantum corrals with Rashba spin$-$orbit coupling\cite{Walls07b}, where multiple scattering trajectories generate noncommuting rotations that result in extra modulations in the LDOS compared to the LDOS found in 2DEGs. Lastly, the LDOS inside the rectangular array is also quite different between graphene and the 2DEG case.  While previous work on a 2DEG has demonstrated that the change in the LDOS can be quite large inside a quantum corral due to quasiparticle interference\cite{Crommie93}, these confinement resonances are, in general, not observed for scattering in graphene corrals, which is a consequence of the inability of the scattering potential to confine the quasiparticles in graphene [from calculations, resonances were not observed in a circular array of $s$-wave scatterers, but some resonances were observed as more partial waves were included in the calculation\cite{Wallstbp}].
  It should be noted that since the scatterers used in the above theory affect both the $A$ and $B$ lattice sites identically, the LDOS calculated from Eq.~(\ref{eq:Gfin}) and shown in Fig.~\ref{fig:rect} would be valid description of  low-resolution/coarse-grained STM images with spatial resolutions larger than than the $C-C$ bond length, $b=1.42$~\AA.

\begin{figure}
\begin{centering}
\includegraphics[scale=0.35]{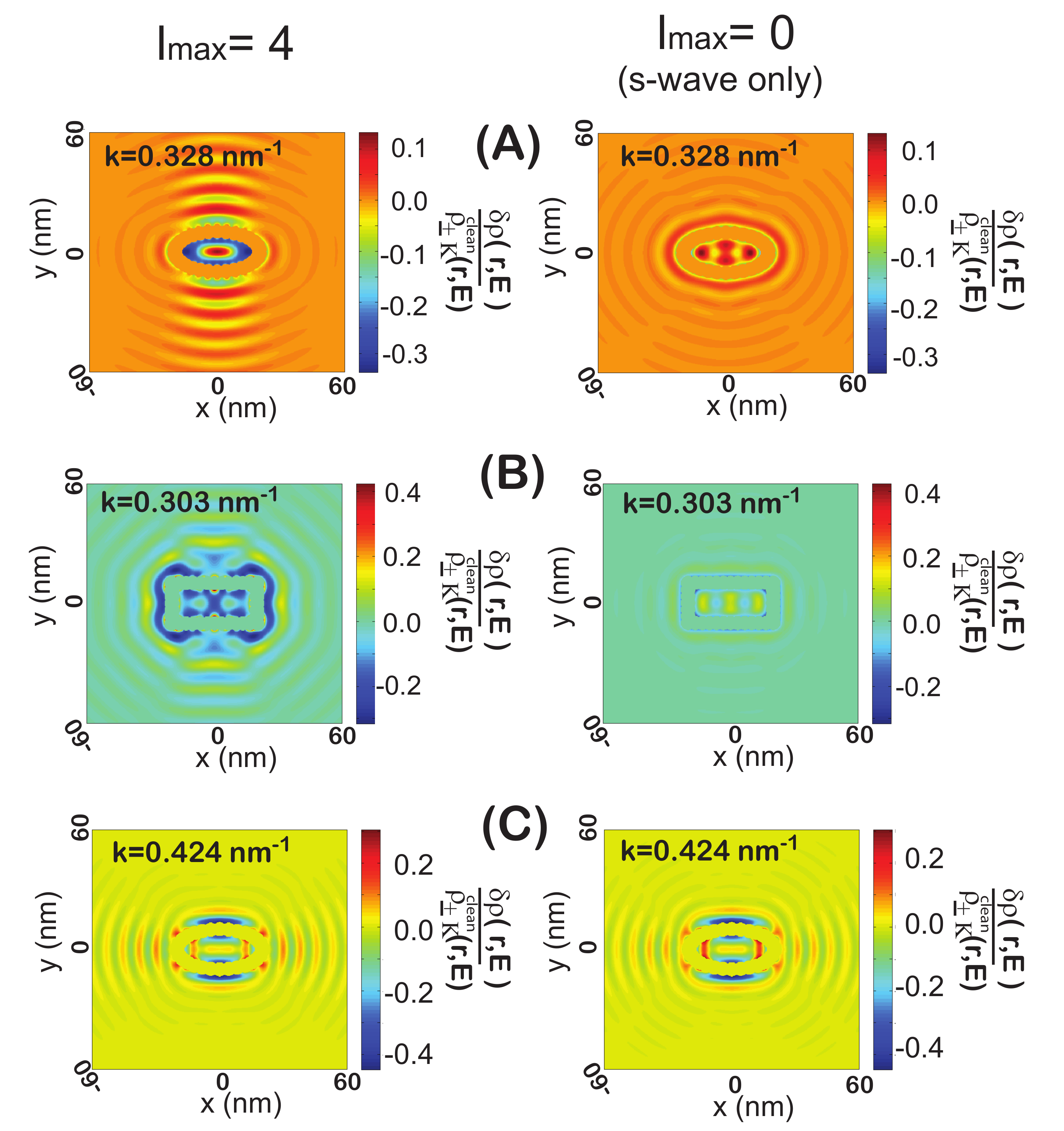}\caption{[Color Online] A comparison of the calculated $\frac{\delta\rho(\vec{r},E)}{\rho^{\text{clean}}_{\pm\vec{K}}(\vec{r},E)}$ [Eq.~(\ref{eq:Gfin})]  using either the first five partial waves [left, $l_{max}=4$] or only the first or $s$-wave scattering [$l_{max}=0$, right] for [Fig. \ref{fig:ellipse}(B)] a 200~\AA~by 400~\AA~ rectangular array of $N=36$ identical scatterers [$V=4.93$ eV, $a=10$~\AA]  and [Figs. \ref{fig:ellipse}(A),\ref{fig:ellipse}(C)] an elliptical array [semimajor and semiminor axis of 200~\AA~ and 100~\AA, respectively] composed of $N=30$ identical scatterers [$a=10$~\AA  and either (A) $V=4.93$ eV or (C) $V=4$ eV].  In all cases, $\frac{\delta\rho(\vec{r},E)}{\rho^{\text{clean}}_{\pm\vec{K}}(\vec{r},E)}$ was set to zero within 35~\AA~of the scatterers, and the choice of $l_{max}=4$ [left] was found to provide a relative accuracy of one percent.  For (A) and (B), the energies were chosen such that the $p$-wave scattering amplitude, $|s_{1}|$, was larger than the $s$-wave scattering amplitude, $|s_{0}|$, which gave (A) $|s_{0}|=0.1429<|s_1|=0.9975$ [$k=0.328~\text{nm}^{-1}$, $E=0.2163$ eV] and (B) $|s_{0}|=0.1237<|s_{1}|=0.9914$ [$k=0.303~\text{nm}^{-1}$, $E=0.2$ eV].  For the elliptical array in (C), however, even though $|s_{0}|=0.9319>|s_{1}|=0.0147$ [$k=0.424~\text{nm}^{-1}$, $E=0.28$ eV], there are still slight differences in $\frac{\delta\rho(\vec{r},E)}{\rho^{\text{clean}}_{\pm\vec{K}}(\vec{r},E)}$ near the array of scatterers between the calculations with $l_{max}=4$ [left] and $l_{max}=0$ [right].           \label{fig:ellipse}}

\par\end{centering}

\end{figure}

In the calculations shown in Fig.~\ref{fig:rect}, $l_{max}=4$ was chosen in order to achieve a relative accuracy in the $\frac{\delta\rho(\vec{r},E)}{\rho^{\text{clean}}_{\pm\vec{K}}(\vec{r},E)}$ profiles of one percent, even though, on average, $|s_0|\gg |s_{l\neq 0}|$ [in Fig.~\ref{fig:rect}, $|s_{0}|\approx 0.94$, $|s_{1}|\approx 1.5\times 10^{-2}|s_{0}|$, $|s_{2}|\approx 6.5\times 10^{-4}|s_{0}|$, $|s_{3}|\approx 3\times 10^{-6}$, and $|s_{4}|\approx 6\times 10^{-9}|s_{0}|$].   However, what matters in the calculation of $\widehat{G}_{\pm\vec{K}}(\vec{r},\vec{r}',E)$ in Eq.~(\ref{eq:Gfin}) is the inverse of $\mathbf{\widehat{\widehat{TT}}}$, which contains off-diagonal terms that are proportional to $H^{(1)}_{l}(kr_{nm})s_{l'}$ that can be quite large when $kr_{nm}\ll 1$.  Therefore, as the distance between scatterers decreases, more partial waves must be included in the calculation.  This is due to the fact that when the distance between scatterers is comparable to their scattering length/size, more partial waves are needed in order to insure that the total wave function satisfies the boundary conditions at the edge of each scatterer [$\rho_{n}=a$].  In Fig.~\ref{fig:rect}, the nearest neighbor distance between scatterers was $33.33$~\AA~$\approx 1.6\times 2a$, which required $l_{max}=4$ to achieve a relative accuracy of one percent.  This is illlustrated in Fig.~\ref{fig:ellipse}(C), where the calculations of the LDOS for an elliptical array [semimajor and semiminor axis of 20 nm and 10 nm respectively] of $N=30$ scatterers [$a=1$ nm and $V=4$ eV] are shown for $l_{max}=4$ [left] and $l_{max}=0$ [right].  In this calculation, $k=0.424\text{nm}^{-1}$ and $|s_{0}|=0.9319\gg |s_{1}|=0.0147$;  however, the inclusion of the first $l_{max}+1=5$ partial waves results in small changes to the LDOS that can be seen around the elliptical array and in the intensity of the Friedel oscillations along the semimajor axis.  For the 2DEG/achiral case, the calculations of the LDOS converged with fewer partial waves, i.e., the results were less sensitive to $l_{max}$ under the same conditions as those used in the graphene calculations.  This is most likely due to the fact that the $l^{th}$ partial wave is composed of both $H^{(1)}_{l}(kr)$ and $H^{(1)}_{l+1}(kr)$, whereas the $l^{th}$ partial wave is composed of  $H^{(1)}_{l}(kr)$ in a 2DEG.  It should also be mentioned that higher partial waves can lead to proximity resonances\cite{Heller96} as the scatterers are placed closer to one another; preliminary calculations indicate that these resonances, which are not observed for $s$-wave scattering only [$l_{max}=0$], lead to large changes in the LDOS of graphene and should be experimentally observable\cite{Wallstbp}.

Finally, it should be noted that the scattering amplitudes in graphene, $s_{l}$ in Eq.~(\ref{eq:phaseshift}), are quite different than those found in 2DEGs for scattering from a radial potential $V(\vec{r})$,  $s_{l,ac}=\frac{J_{l}(k'a)}{H^{(1)}_{l}(ka)}$.  These differences are attributable to the chiral nature of quasiparticles in graphene, where the wave function of both $A$ and $B$ lattice sites must be matched at each scatterer boundary.  One consequence is that there exist conditions where $|s_{1}|>|s_{0}|$ even though $ka<1$, which is not true for $s_{l,ac}$.  In these cases, taking into account only $s$-wave scattering can lead to large errors in the calculated LDOS.  This is illustrated for both an elliptical array [Fig.~\ref{fig:ellipse}(A)] and a rectangular array [Fig.~\ref{fig:ellipse}(B)], where the parameters used in the simulation are given in the caption of Fig.~\ref{fig:ellipse}.  In both cases, $|s_{1}|\approx 7|s_{0}|$, which leads to large differences in the calculated LDOS if only $s$-wave scattering is considered;  this is clearly illustrated in Figs. \ref{fig:ellipse}(A) and \ref{fig:ellipse}(B).

\section{Conclusions}
In this work, a theory for intravalley multiple scattering in single layer graphene was developed and used to calculate the total scattered wave function, Green's function, and the change in the local density of states (LDOS) in the presence of multiple scatterers.  The scatterers were modeled by electrostatic step potentials that either raise or lower the potential within a radius $a$, with $a$ being much larger than the $C-C$ bond length, $b=1.42$~\AA.  Such scatterers could be experimentally realized by placing small metallic islands atop a graphene surface\cite{Titov10,Kessler10}.  In graphene, the shape, intensity, and pattern of quasiparticle interference were shown to be affected by the chiral nature of graphene compared with simple two-dimensional electron gases (2DEGs) or 2D achiral systems.  The effects of these pseudospin rotations on the local density of states (LDOS) in graphene are similar to those predicted for the case of spin$-$orbit coupling in 2DEGS\cite{Walls07b}.  Furthermore, the LDOS was found to be sensitive to the inclusion of higher partial waves if the distance between nearby scatterers was close to 2$a$, even in the $s$-wave scattering scattering limit [i.e., $|s_{0}|\gg |s_{l\neq 0}|$ in Eq. (\ref{eq:phaseshift})].  The theoretical predictions in this work should be verifiable using STM\cite{Rutter07,Brihuega08} and in atomic\cite{hexagonatom} or microwave/optical\cite{Bittner10} experimental ``simulations" of graphene systems.  Finally, since the total Green's function is given by Eq.~(\ref{eq:Gfin}), the theory presented in this work could be used to calculate the effects of multiple scattering on the transport properties in graphene under conditions where intervalley scattering are negligible.

The ultimate goal of this work was to lay out a formalism that enables the relatively fast calculation of the Green's function in single layer graphene for arbitrary scattering configurations and that elucidates the role that pseudospin interference has on the LDOS in graphene.  A clear understanding of impurity scattering in graphene provides a step toward someday exploiting graphene's unique properties to build new electronic devices.  In the future, the theory in this work could be extended to scattering in multiple-layer graphene, nonzero bandgap graphene monolayers, higher energies, and to intervalley scattering.  More realistic models for the scatterers could also be considered, where only the individual scattering amplitudes/phase shifts are needed as input in Eq.~(\ref{eq:Gfin}).   Finally, resonances corresponding to those predicted for neutrino billiards might also be observable\cite{Berry87} for scatterers that locally induce a nonzero bandgap.

\section{Acknowledgments}
We would like to thank Dr. John W. Logan for comments on the manuscript.  JYV and JQA wish to thank Dr. Charles W. Clark for his support.  JDW was supported by a Camille and Henry Dreyfus New faculty award and startup funds from the University of Miami.

\appendix
\section{LDOS for 2DEG/achiral system}
The $t-$matrix for scattering of higher partial waves by point scatterers in a 2DEG/2D achiral system has been previously derived by Hersch\cite{Hersch99}.  In this case, the total Green's function in the presence of $N$ scatterers is given by:
\begin{eqnarray}
\widehat{G}(\vec{r},\vec{r}',E)&=&\widehat{G}_{0}(\vec{r},\vec{r}',E)+\sum_{j=1}^{N}\frac{2i\hbar^{2}s^{(j)}_{0}}{m}\widehat{G}_{0}(\vec{r},\vec{r}_{j},E)\widehat{G}(\vec{r}_{j},\vec{r}',E)\nonumber\\
&+&\sum_{j=1}^{N}\sum_{l=1}^{l_{max}}\frac{2i\hbar^{2}s^{(j)}_{l}}{m}\left(\hat{L}^{l}_{+}\left[\widehat{G}_{0}(\vec{r},\vec{r}_{j},E)\right]\hat{L}^{l}_{-}\left[\widehat{G}_{0}(\vec{r}_{j},\vec{r}',E)\right]+\hat{L}^{l}_{-}\left[\widehat{G}_{0}(\vec{r},\vec{r}_{j},E)\right]\hat{L}^{l}_{+}\left[\widehat{G}(\vec{r}_{j},\vec{r}',E)\right]\right)\nonumber\\
&=&-i\frac{m}{2\hbar^{2}}H^{(1)}_{0}(k|\vec{r}-\vec{r}'|)+\sum_{j=1}^{N}s^{(j)}_{0}H^{(1)}_{0}(k\rho_{j})\widehat{G}(\vec{r}_{j},\vec{r}',E)\nonumber\\
&+&\sum_{j=1}^{N}\sum_{l=1}^{l_{max}}s_{l}^{(j)}\left(i^{l}H^{(1)}_{l}(k\rho_{j})e^{il\theta_{j}}\hat{L}^{l}_{-}\left[\widehat{G}(\vec{r}_{j},\vec{r}',E)\right]+i^{-l}H^{(1)}_{-l}(k\rho_{j})e^{-il\theta_{j}}\hat{L}^{l}_{+}\left[\widehat{G}(\vec{r}_{j},\vec{r}',E)\right]\right)
\label{eq:achiralgreeno}
\end{eqnarray}
where $\widehat{G}_{0}(\vec{r},\vec{r}',E)=-i\frac{m}{2\hbar^{2}}H^{(1)}_{0}(k|\vec{r}-\vec{r}'|)$, $\rho_{j}=|\vec{r}-\vec{r}_{j}|$, $e^{il\theta_{j}}=\left(\frac{(\vec{r}-\vec{r}_{j})\cdot\left(\widehat{x}+i\widehat{y}\right)}{\rho_{j}}\right)^{l}$, $k=\sqrt{\frac{2mE}{\hbar^{2}}}$ and $m$ is the mass of the scattered particle.

In this case, there are $N(2l_{max}+1)$ unknowns, which can be solved for self-consistently using the following Foldy$-$Lax equations [for $l=0$ to $l=l_{max}$ and $n=1$ to $n=N$]:
\begin{eqnarray}
\hat{L}^{l}_{\pm}\left[\widehat{G}(\vec{r}_{n},\vec{r}',E)\right]&=&\frac{m}{2\hbar^{2}}\left(-i\right)^{1\pm l}H_{\pm l}(k\rho_{n})e^{\pm il\theta_{j}}+\sum_{j\neq n}s^{(j)}_{0}i^{\pm l}H^{(1)}_{\pm l}(kr_{nj})e^{i\pm l\theta_{nj}}\widehat{G}(\vec{r}_{j},\vec{r}',E)\nonumber\\
&+&\sum_{j\neq n}\sum_{l'=1}^{l_{max}}s^{(j)}_{l'}\left((i)^{l'\pm l}H^{(1)}_{l'\pm l}(kr_{nj})e^{i(l'\pm l)\theta_{nj}}\hat{L}_{-}^{l'}\left[\widehat{G}(\vec{r}_{j},\vec{r}',E)\right]+(i)^{-l'\pm l}H^{(1)}_{-l'\pm l}(kr_{nj})e^{i(-l'\pm l)\theta_{nj}}\hat{L}^{l'}_{+}\left[\widehat{G}(\vec{r}_{j},\vec{r}',E)\right]\right)\nonumber\\
\label{eq:achiralgreen}
\end{eqnarray}
Defining the following $2l_{max}+1$ $\times 1$ vectors for each scatterer $j=1$ to $j=N$:
\begin{eqnarray}
\widehat{\widehat{{T}}}\widehat{\widehat{G}}^{(j)}(\vec{r}_{j},\vec{r}',E)&=&\left(\begin{array}{c}\widehat{G}(\vec{r}_{j},\vec{r}',E)\\
\hat{L}_{+}[\widehat{G}(\vec{r}_{j},\vec{r}',E)]\\
\hat{L}_{-}[\widehat{G}(\vec{r}_{j},\vec{r}',E)]\\
\hat{L}^{2}_{+}[\widehat{G}(\vec{r}_{j},\vec{r}',E)]\\
\hat{L}^{2}_{-}[\widehat{G}(\vec{r}_{j},\vec{r}',E)]\\
\vdots\\
\hat{L}^{l_{max}}_{+}[\widehat{G}(\vec{r}_{j},\vec{r}',E)]\\
\hat{L}^{l_{max}}_{-}[\widehat{G}(\vec{r}_{j},\vec{r}',E)]
\end{array}\right),\widehat{\widehat{T}}\widehat{\widehat{G}}^{(j)}_{0}(\vec{r}_{j},\vec{r}')=-i\frac{m}{2\hbar^{2}}\left(\begin{array}{c}H^{(1)}_{0}(k\rho'_{j})\\
-iH^{(1)}_{1}(k\rho'_{j})e^{i\theta'_{j}}\\
(-i)^{-1}H^{(1)}_{-1}(k\rho'_{j})e^{-i\theta'_{j}}\\
(-i)^{2}H^{(1)}_{2}(k\rho'_{j})e^{i2\theta'_{j}}\\
(-i)^{-2}H^{(1)}_{-2}(k\rho'_{j})e^{-i2\theta'_{j}}\\
\vdots\\
(-i)^{l_{max}}H^{(1)}_{l_{max}}(k\rho'_{j})e^{il_{max}\theta'_{j}}\\
(-i)^{-l_{max}}H^{(1)}_{-l_{max}}(k\rho'_{j})e^{-il_{max}\theta'_{j}}
\end{array}\right)
\end{eqnarray}
where $\rho'_{j}=|\vec{r}'-\vec{r}_{j}|$ and $e^{\pm i\theta'_{j}}=\frac{(\vec{r}'-\vec{r}_{j})\cdot(\widehat{x}\pm i\widehat{y})}{\rho'_{j}}$.  The following $N(2l_{max}+1)\times 1$ column vectors can then be constructed:
\begin{eqnarray}
\mathbf{\widehat{\widehat{T}}}_{ac}\mathbf{\widehat{\widehat{G}}}(\vec{r}')=\left(\begin{array}{c}
\widehat{\widehat{T}}\widehat{\widehat{G}}^{(1)}(\vec{r}_{1},\vec{r}')\\
\widehat{\widehat{T}}\widehat{\widehat{G}}^{(2)}(\vec{r}_{2},\vec{r}')\\
\vdots\\
\widehat{\widehat{T}}\widehat{\widehat{G}}^{(N)}(\vec{r}_{N},\vec{r}')
\end{array}\right),
\mathbf{\widehat{\widehat{T}}}_{ac}\mathbf{\widehat{\widehat{G}}}_{0}(\vec{r}')=\left(\begin{array}{c}
\widehat{\widehat{T}}\widehat{\widehat{G}}^{(1)}_{0}(\vec{r}_{1},\vec{r}')\\
\widehat{\widehat{T}}\widehat{\widehat{G}}^{(2)}_{0}(\vec{r}_{2},\vec{r}')\\
\vdots\\
\widehat{\widehat{T}}\widehat{\widehat{G}}^{(N)}_{0}(\vec{r}_{N},\vec{r}')
\end{array}\right)
\end{eqnarray}
Eqs. (\ref{eq:achiralgreen}) can be written compactly for $n=1$ to $n=N$ and for $l=0$ to $l=l_{max}$ as:
\begin{eqnarray}
\mathbf{\widehat{\widehat{T}}}_{ac}\mathbf{\widehat{\widehat{G}}}(\vec{r}')&=&\left(\mathbf{\widehat{\widehat{1}}}-\mathbf{\widehat{\widehat{TT}}_{ac}}\right)^{-1}\mathbf{\widehat{\widehat{T}}}_{ac}\mathbf{\widehat{\widehat{G}}}_{0}(\vec{r}')
\end{eqnarray}
where
$\mathbf{\widehat{\widehat{1}}}$ is the $N(2l_{max}+1)\times N(2l_{max}+1)$ identity matrix and $\mathbf{\widehat{\widehat{TT}}_{ac}}$ is a $N(2l_{max}+1)\times N(2l_{max}+1)$ matrix given by:
\begin{eqnarray}
\mathbf{\widehat{\widehat{TT}}_{ac}}&=&\left(\begin{array}{ccccc}\mathbf{\widehat{0}}&\widehat{\widehat{TT}}_{ac}(\vec{r}_{1},\vec{r}_{2})&
\widehat{\widehat{TT}}_{ac}(\vec{r}_{1},\vec{r}_{3})&\hdots&\widehat{\widehat{TT}}_{ac}(\vec{r}_{1},\vec{r}_{N})\\
\widehat{\widehat{TT}}_{ac}(\vec{r}_{2},\vec{r}_{1})&\mathbf{\widehat{0}}&\widehat{\widehat{TT}}_{ac}(\vec{r}_{2},\vec{r}_{3})&\hdots&\widehat{\widehat{TT}}_{ac}(\vec{r}_{2},\vec{r}_{N})\\
\vdots&\vdots&\vdots&\ddots&\vdots\\
\widehat{\widehat{TT}}_{ac}(\vec{r}_{N},\vec{r}_{1})&\widehat{\widehat{TT}}_{ac}(\vec{r}_{N},\vec{r}_{2})&\widehat{\widehat{TT}}_{ac}(\vec{r}_{N},\vec{r}_{3})&\hdots&\mathbf{\widehat{0}}
\end{array}
\right)
\end{eqnarray}
where {\small{\begin{eqnarray}
\widehat{\widehat{TT}}_{ac}(\vec{r}_{n},\vec{r}_{j})&=&\left(\begin{array}{cccccc}s^{(j)}_{0}H^{(1)}_{0}(kr_{nj})&s^{(j)}_{1}\hat{L}_{-}\left[H_{0}^{(1)}(kr_{nj})\right]&s^{(j)}_{1}\hat{L}_{+}\left[H_{0}^{(1)}(kr_{nj})\right]&\hdots&s^{(j)}_{l_{max}}\hat{L}_{-}^{l_{max}}\left[H_{0}^{(1)}(kr_{nj})\right]&s^{(j)}_{l_{max}}\hat{L}_{+}^{l_{max}}\left[H_{0}^{(1)}(kr_{nj})\right]\\
s^{(j)}_{0}\hat{L}_{+}\left[H^{(1)}_{0}(kr_{nj})\right]&s^{(j)}_{1}H_{0}^{(1)}(kr_{nj})&s^{(j)}_{1}\hat{L}^{2}_{+}\left[H_{0}^{(1)}(kr_{nj})\right]&\hdots&s^{(j)}_{l_{max}}\hat{L}_{-}^{l_{max}-1}\left[H_{0}^{(1)}(kr_{nj})\right]&s^{(j)}_{l_{max}}\hat{L}_{+}^{l_{max}+1}\left[H_{0}^{(1)}(kr_{nj})\right]\\
s^{(j)}_{0}\hat{L}_{-}\left[H^{(1)}_{0}(kr_{nj})\right]&s^{(j)}_{1}\hat{L}^{2}_{-}\left[H_{0}^{(1)}(kr_{nj})\right]&s^{(j)}_{1}H_{0}^{(1)}(kr_{nj})&\hdots&s^{(j)}_{l_{max}}\hat{L}_{-}^{l_{max}+1}\left[H_{0}^{(1)}(kr_{nj})\right]&s^{(j)}_{l_{max}}\hat{L}_{+}^{l_{max}-1}\left[H_{0}^{(1)}(kr_{nj})\right]\\
\vdots&\vdots&\vdots&\ddots&\vdots&\vdots\\
s^{(j)}_{0}\hat{L}^{l_{max}}_{+}\left[H^{(1)}_{0}(kr_{nj})\right]&s^{(j)}_{1}\hat{L}^{l_{max}-1}_{+}\left[H_{0}^{(1)}(kr_{nj})\right]&s^{(j)}_{1}\hat{L}^{l_{max}+1}_{+}\left[H_{0}^{(1)}(kr_{nj})\right]&\hdots&s^{(j)}_{l_{max}}H_{0}^{(1)}(kr_{nj})&s^{(j)}_{l_{max}}\hat{L}_{+}^{2l_{max}}\left[H_{0}^{(1)}(kr_{nj})\right]\\
s^{(j)}_{0}\hat{L}^{l_{max}}_{-}\left[H^{(1)}_{0}(kr_{nj})\right]&s^{(j)}_{1}\hat{L}^{l_{max}+1}_{-}\left[H_{0}^{(1)}(kr_{nj})\right]&s^{(j)}_{1}\hat{L}^{l_{max}-1}_{-}\left[H_{0}^{(1)}(kr_{nj})\right]&\hdots&s^{(j)}_{l_{max}}\hat{L}^{2l_{max}}_{-}\left[H_{0}^{(1)}(kr_{nj})\right]&s^{(j)}_{l_{max}}H_{0}^{(1)}(kr_{nj})
\end{array}\right)\nonumber\\
\end{eqnarray}}}
Finally, $\widehat{G}(\vec{r},\vec{r}',E)$ in Eq.~(\ref{eq:achiralgreeno}) can be written compactly as:
\begin{eqnarray}
\widehat{G}(\vec{r},\vec{r}',E)&=&-i\frac{m}{2\hbar^{2}}H^{(1)}_{0}(k|\vec{r}-\vec{r}'|)+\mathbf{\widehat{\widehat{GG}}_{ac}}(\vec{r})\mathbf{\widehat{\widehat{T}}_{ac}\widehat{\widehat{G}}}_{\pm\vec{K}}(\vec{r}')\nonumber\\
&=&-i\frac{m}{2\hbar^{2}}H^{(1)}_{0}(k|\vec{r}-\vec{r}'|)+\mathbf{\widehat{\widehat{GG}}_{ac}}(\vec{r})\left(\mathbf{\widehat{\widehat{1}}}-\mathbf{\widehat{\widehat{TT}}_{ac}}\right)^{-1}\mathbf{\widehat{\widehat{T}}_{ac}\widehat{\widehat{G}}}_{0,\pm\vec{K}}(\vec{r}')
\label{eq:Gfinac}
\end{eqnarray}
where $\mathbf{\widehat{\widehat{GG}}_{ac}}(\vec{r})$ is a $1\times N(2l_{max}+1)$ vector, $\mathbf{\widehat{\widehat{GG}}_{ac}}(\vec{r})=\left[\widehat{\widehat{GG}}_{ac}(\vec{r},\vec{r}_{1})\,\,\widehat{\widehat{GG}}_{ac}(\vec{r},\vec{r}_{2})\,\,\hdots\widehat{\widehat{GG}}_{ac}(\vec{r},\vec{r}_{N})\right]$
where {\small{
\begin{eqnarray}
\widehat{\widehat{GG}}_{ac}(\vec{r},\vec{r}_{j})&=&\left[\begin{array}{cccccc}s^{(j)}_{0}H^{(1)}_{0}(k\rho_{j}),&s^{(j)}_{1}i^{-1}H^{(1)}_{-1}(k\rho_{j})e^{-i\theta_{j}},&s^{(j)}_{1}iH^{(1)}_{1}(k\rho_{j})e^{i\theta_{j}},&\hdots&s^{(j)}_{l_{max}}i^{-l_{max}}H^{(1)}_{-l_{max}}(k\rho_{j})e^{-il_{max}\theta_{j}},&s^{(j)}_{l_{max}}i^{l_{max}}H^{(1)}_{l_{max}}(k\rho_{j})e^{il_{max}\theta_{j}}\end{array}\right]\nonumber\\
\end{eqnarray}}}
The 2DEG Green's function, $\widehat{G}(\vec{r},\vec{r}',E)$ in Eq.~(\ref{eq:Gfinac}), was used to calculate $\frac{\delta\rho(\vec{r},E)}{\rho^{\text{clean}}(\vec{r},E)}$ in Fig.~\ref{fig:figure2} and Fig.~\ref{fig:rect} for the 2DEG/achiral cases.
\section{Equivalence between Foldy$-$Lax method and the point scatterer model}

In condensed matter systems, scatterers are often modeled as simple point scatterers.  In this model, the scatterers are described by a sum of effecitve $\delta$-potentials, $V(\vec{r})=\sum_{j=1}^{N}V_{j}\delta(\vec{r}-\vec{r}_{j})$ where $V_{j}$, which has units of energy$\times$area, is the potential associated with scatterer $j$.  In this case, the total Green's function in the presence of $N$ point scatterers, which we denote by $\widehat{G}^{s}_{\pm\vec{K}}(\vec{r},\vec{r}',E)$, is given by the Lippmann-Schwinger equation:
\begin{eqnarray}
\widehat{G}^{s}_{\pm\vec{K}}(\vec{r},\vec{r}',E)&=&\widehat{G}_{0,\pm\vec{K}}(\vec{r},\vec{r}',E)+\int\text{d}^{2}\vec{r}''\widehat{G}_{0,\pm\vec{K}}(\vec{r},\vec{r}'',E)V(\vec{r}'')\widehat{G}^{s}_{\pm\vec{K}}(\vec{r}'',\vec{r}',E)\nonumber\\
\widehat{G}^{s}_{\pm\vec{K}}(\vec{r},\vec{r}',E)&=&\widehat{G}_{0,\pm\vec{K}}(\vec{r},\vec{r}',E)+\sum_{j=1}^{N}V_{j}\widehat{G}_{0,\pm\vec{K}}(\vec{r},\vec{r}_{j},E)\widehat{G}^{s}_{\pm\vec{K}}(\vec{r}_{j},\vec{r}',E)
\label{eq:greenso}
\end{eqnarray}
As in the previous section, the various $\widehat{G}^{s}_{\pm\vec{K}}(\vec{r}_{j},\vec{r}',E)$ can be solved for self-consistently by the following equations for $j=1$ to $j=N$:
\begin{eqnarray}
\widehat{G}^{s}_{\pm\vec{K}}(\vec{r}_{j},\vec{r}',E)&=&\widehat{G}_{0,\pm\vec{K}}(\vec{r}_{j},\vec{r}',E)+V_{j}\overline{G}_{0,\pm\vec{K}}(\vec{r}_{j},\vec{r}_{j},E)\widehat{G}^{s}_{\pm\vec{K}}(\vec{r}_{j},\vec{r}',E)+\sum_{n\neq j}V_{n}\widehat{G}_{0,\pm\vec{K}}(\vec{r}_{j},\vec{r}_{n},E)\widehat{G}^{s}_{\pm\vec{K}}(\vec{r}_{n},\vec{r}',E)
\label{eq:greensself}
\end{eqnarray}
Unlike the self-consistent Foldy$-$Lax\cite{Foldy45,Lax51} equations in Eq.~(\ref{eq:selfgreen}), $\widehat{G}^{s}_{\pm\vec{K}}(\vec{r}_{j},\vec{r}',E)$ in Eq.~(\ref{eq:greensself}) is scaled by $(\widehat{1}-\overline{G}_{0,\pm\vec{K}}(\vec{r}_{j},\vec{r}_{j},E))$, where $\overline{G}_{0,\pm\vec{K}}(\vec{r}_{j},\vec{r}_{j},E)$ is the ``renormalized" Green's function of $\widehat{G}_{0,\pm\vec{K}}(\vec{r},\vec{r}',E)$ at $\vec{r}=\vec{r}'=\vec{r}_{j}$ where the logarithmic singularity has been removed, i.e., $\overline{G}_{0,\pm\vec{K}}(\vec{r}_{j},\vec{r}_{j},E)=-\frac{ik}{4\hbar\nu_{F}}\left(1+i\frac{2}{\pi}\left[\text{ln}\left(\frac{kd}{2}\right)+\gamma\right]\right)\left(\begin{array}{cc}1&0\\0&1\end{array}\right)_{\pm\vec{K}}$, where $d$ is a renormalization length, and $\gamma=0.5772$ is the Euler-Mascheroni constant.  The solution to Eqs. (\ref{eq:greensself}) for the values of $\widehat{G}_{\pm\vec{K}}(\vec{r}_{j},\vec{r}',E)$ for $j=1$ to $j=N$ and therefore the total Green's function, $\widehat{G}^{s}_{\pm\vec{K}}(\vec{r},\vec{r}',E)$, can be written compactly using the Foldy$-$Lax eqatuion for $s$-wave scattering [Eq.~(\ref{eq:Gfin}) with $l_{max}=0$] but with the $s$-wave scattering amplitude for the $j^{th}$ scatterer, $s^{(j)}_{0}$, replaced by
\begin{eqnarray}
\tilde{s}^{(j)}_{0}&=&\frac{-i\frac{kV_{j}}{4\hbar\nu_{F}}}{\left[1-\frac{kV_{j}}{2\pi\hbar\nu_{F}}\left(\ln\left(\frac{kd}{2}\right)+\gamma\right)\right]+i\frac{kV_{j}}{4\hbar\nu_{F}}}
\label{eq:spot}
\end{eqnarray}
It is easy to verify that the point scatterer scattering amplitude, $\tilde{s}^{(j)}_{0}$ in Eq.~(\ref{eq:spot}), satisfies the unitarity condition, Re$[\tilde{s}^{(j)}_{0}]=-|\tilde{s}^{(j)}_{0}|^{2}$.   Thus the point-scatterer model generates the same Green's function found using the Foldy$-$Lax equations for $s$-wave scattering but with the $s$-wave scattering amplitude given by Eq.~(\ref{eq:spot}).

%{\bibliographystyle{prsty}
%\bibliography{wallsbib}

\end{document}